\documentclass[aps,prd,preprint]{revtex4-1}

\usepackage{latexsym}
\usepackage{tensor}
\usepackage{amsmath}
\usepackage{amsfonts}
\usepackage{amssymb}

\allowdisplaybreaks

\usepackage{graphicx}

\makeatletter
\DeclareRobustCommand{\cev}[1]{%
  {\mathpalette\do@cev{#1}}%
}
\newcommand{\do@cev}[2]{%
  \vbox{\offinterlineskip
    \sbox\z@{$\m@th#1 x$}%
    \ialign{##\cr
      \hidewidth\reflectbox{$\m@th#1\vec{}\mkern4mu$}\hidewidth\cr
      \noalign{\kern-\ht\z@}
      $\m@th#1#2$\cr
    } 
  } 
}

\usepackage{appendix} 

\fontsize{12}{1} \selectfont
\usepackage[indentafter]{titlesec}

\begin{document}

\title{The Graded Extension of Thomas$-$Whitehead  Gravity}

\author{Calvin Mera S\'anchez}
 \affiliation{%
 Department of Physics and Astronomy\\
 The University of Iowa, Iowa City, IA 52242, USA}
\email{calvin-mera@uiowa.edu}\email{vincent-rodgers@uiowa.edu}

 \author{Vincent G. J. Rodgers}
 \affiliation{%
 Department of Physics and Astronomy\\
 The University of Iowa, Iowa City, IA 52242, USA}
 \email{vincent-rodgers@uiowa.edu}

\author{Patrick Vecera}
\affiliation{%
Department of Mathematics\\
University of California
Santa Barbara, CA 93106-3080}
\email{pvecera@ucsb.edu}

\date{\today}

\begin{abstract}
Thomas-Whitehead (TW) gravity was recently introduced as a  projective gauge theory of gravity over a d-dimensional manifold that embeds reparameterization invariance into the action functional for gravitation through the use of the Thomas-Whitehead connection.  The projective invariance in this d-dimensional theory enjoys an intimate relationship with the Virasoro coadjoint elements found in string theory as one of the components of the connection,  $\mathcal{D}_{ab}$, is directly related to the coadjoint elements  of the Virasoro algebra.  TW Gravity exploits projective Gauss-Bonnet terms in the action functional  which allows the theory to collapse to Einstein's theory of General Relativity in the limit that  $\mathcal{D}_{ab }$ vanishes.   In this note we develop the graded extension of  TW Gravity, Super TW Gravity,  in the framework of a DeWitt supermanifold. We construct the Lagrangian for  Super TW Gravity, give a detailed derivation of  the classical field equations and discuss the graded extension of the projective connection as a prelude to a future understanding of TW-Supergravity (which has manifest supersymmetry) and its relationship to the Super-Virasoro algebra.     
\end{abstract}

\maketitle

%% arabic for the rest
\pagenumbering{arabic}

%\begin{onehalfspacing}
%% main contents %%
\section{Introduction}

It is known that the method of coadjoint orbits \cite{Kirillov:1962,Kirillov:1982kav} of the semi-direct product of Kac-Moody algebras and the Virasoro algebras \cite{Rai:1989js,Alekseev:1988ce,Alekseev:1988vx,Delius:1990pt} leads  to the two-dimensional Wess-Zumino-Witten (WZW) action  \cite{DiVecchia:1984ksr,Witten:1983ar} and the Polyakov action  \cite{Polyakov:1987zb,Polyakov:1981rd} respectively. One may arrive at this by integrating  the the Kirillov two-form  \cite{Kirillov:1962,Kirillov:1982kav} over any coadjoint orbit as prescribed in \cite{Balachandran:1979pc,Balachandran:1986hv,Balachandran:1987st} which produces these \emph{geometric} actions for their respective  groups. One finds that the coadjoint elements have been promoted to fields in the geometric actions and the central extension to a coupling constant. The geometric actions  interprets the elements  of the coadjoint representation of the Virasoro algebra   as a background fields coupling to the Polyakov metric. For the Virasoro Group this  background field been called the diffeomorphism field $\mathcal D_{\mu \nu}$ and is akin to the Yang-Mills connection, $  A_\mu$,  that related the coadjoint elements of the  Kac-Moody Group.    One can extend these  geometric actions by adding dynamics to  $A_\mu$ through the addition of  the Yang-Mills action in the WZW case, and similarly by adding the  Thomas-Whitehead (TW) action \cite{Brensinger:2017gtb,Brensinger:2020gcv,Brensinger:2019mnx} to the Polyakov action to give dynamics to the $\mathcal D_{\mu \nu}$. This  reconciles the coadjoint elements of both the Kac-Moody algebra and the Virasoro algebra with geometric connections in higher dimensions. 
In \cite{Brensinger:2020gcv} a detailed overview of  Thomas-Whitehead  gravity in a general setting is discussed.  This includes a review of  the relationship between the projective structure from Sturm-Liouville theory and the two-cocycles of the Virasoro algebra as observed by Kirillov \cite{Kirillov:1982kav,OvsienkoValentin2005Pdgo}  as well as a derivation of the spin connection on the Thomas Cone \cite{Eastwood2}, the  Dirac equation   and the Dirac Lagrangian for spin $\frac{1}{2}$ spinors (fermions).

In \cite{Delius:1990pt,Aoyama:1989pw,Bershadsky:1989tc}, the authors  applied the method of coadjoint orbits to the super Virasoro algebra and later extended \cite{Gates:2001uu,Gates:2002xh} in the context of studying superstring theories. This  recovered the supersymmetric extension of Polyakov's action. A natural question to ask is what the supersymmetric extension of TW gravity is. In this note, we study the preliminary  question by generalizing the theory of Thomas-Whitehead gravity to a supermanifold with $n$ ordinary coordinates and $m$ Grassmann coordinates.
A highly detailed discussion of the calculations can be found in \cite{Mera2020}.   

In chapter II we briefly  review the theory of TW Gravity and in chapter III briefly describe supervector spaces and supermanifolds  following the approach of \cite{Gates:2001uu,DeWitt:2012mdz}. The consequences of the DeWitt topology and its relation to the theory of supersymmetry is reviewed from the perspective of \cite{Rogers:1979vp,Rogers:2007zza}. Other excellent references on the theory of supermanifolds include \cite{Kostant:1975qe, Rogers:2007zza,Berezin:1987wh,Leites:1980rna}.  In chapter IV we generalize the TW connection to the graded manifolds by following the approach in \cite{Brensinger:2020gcv}. The theory of TW connections has been studied in the graded setting before in \cite{JGeorge,Leuther:2010zv,Leuther2013GeodesicsOA}.

\section{Review of Thomas-Whitehead Gravity}
In this section we review the TW Gravity as developed in \cite{Brensinger:2017gtb,Brensinger:2019mnx,Brensinger:2020gcv}.  Projective geometry, as a theory of gravity, has been around for nearly a century as a strategy to incorporate the ambiguity of  geodesics in relation to  connections due to projective transformations \cite{Veblen1930,Thomas:1925a,Thomas:1925b}.   The Thomas-Whitehead Gravitational action \cite{Brensinger:2017gtb} (named after mathematicians Tracy Thomas and J.H.C. Whitehead) uses the covariant derivative and the fibration of the Thomas Cone from these early investigators to tie projective geometry to string theory and higher dimensional gravity through projective Gauss-Bennet terms on the manifold. The projective connection and the metric are treated as independent in the spirit of the Palatini formalism \cite{palatini}.   This allows the field equations  to collapse naturally to the Einstein-Hilbert field equations when the diff field  vanishes and when the fundamental projective invariant is evaluated on the affine connection compatible with the Einstein metric. In this way projective geometry can influence the  Riemannian geometry by acting as sources in the energy-momentum tensor.  This provides an avenue for  geometric explanations of dark energy, dark matter, and other physical phenomenon. 

To proceed we are given a connection    $\tensor{\Gamma}{^a_{bc}}$ on a Riemannian manifold. One can define the fundamental projective invariant as 
\begin{equation}
\tensor{\Pi}{^a_{bc}}\equiv\tensor{\Gamma}{^a_{bc}}-\frac{1}{m+1}\big(\tensor{\Gamma}{^d_{dc}}\tensor{\delta}{^a_b}+\tensor{\Gamma}{^d_{db}}\tensor{\delta}{^a_c}\big),
\end{equation}
which is invariant under projective transformations 
\begin{align}
\tensor{\hat{\Gamma}}{^a_{bc}} = \Gamma^a_{\ bc} + \delta^a_{\ b} v_c + \delta^a_{\ c} v_b.   
\end{align}
Let $J$ be the Jacobian of the coordinate transformation $x^a\rightarrow y^a$. Then we have the following identities: 
\begin{align}
\partial_c\text{log}(\text{det}(J)) & = - \tensor{J}{^a_b} \partial_c \tensor{(J^{-1})}{^b_a} = - \tensor{J}{^a_b} \partial_c \tensor{J}{_a^b} , \\
\tensor{J}{^h_f} \frac{\partial^2 x^f}{\partial y^h\partial y^c} &  = - \partial_m \text{log}(\text{det}(J)) \tensor{(J^{-1})}{^m_c}.
\end{align}
With this,  the coordinate transformation law of the fundamental projective invariant is 
\begin{align}
\label{eq: Pi transformation}
\tensor{\bar{\Pi}}{^a_{bc}}
&=\tensor{J}{^a_f}\bigg(\tensor{\Pi}{^f_{de}}\tensor{(J^{-1})}{^e_c}\tensor{(J^{-1})}{^d_b}+\frac{\partial^2 x^f}{\partial y^b\partial y^c}\bigg)\\
&+\frac{1}{m+1}\frac{\partial}{\partial x^m}\text{log}(\text{det}(J))\big(\tensor{(J^{-1})}{^m_c}\tensor{\delta}{^a_b}+\tensor{(J^{-1})}{^m_b}\tensor{\delta}{^a_c}\big).\nonumber
\end{align}

From equation \ref{eq: Pi transformation} it is apparent that $\Pi$ itself is not a connection due to the extra terms arising in the transformation law. To construct a connection realizing projective invariance we adopt the approach of Thomas \cite{Thomas:1925a,Thomas:1925b} and  consider a connection not on $M$ but instead on the volume bundle $VM$, which is now called the \emph{Thomas Cone}. 

The Thomas cone arises as follows. Given an $m-$manifold $M$, a volume form can be constructed from a smooth nonvanishing function $v:M\rightarrow \mathbb{R}_+$ and considering the $m$ - form 
\begin{align}
|v(x)| \ dx^1 \wedge...\wedge dx^m, 
\end{align}
which is a section of the volume bundle $VM$. We take the absolute value of $v$ to absolve the ambiguity of choice of orientation. $VM$ is then defined as the collection of all such sections, and is an $\mathbb{R}^+$ line bundle over $M$. As a manifold, $VM$ is one dimension higher than $M$.   

We use $\lambda$ as the fiber coordinate on the Thomas Cone, so the coordinates on $VM$ are $(x^0, x^1,...,x^{m-1}, \lambda)$, where $0 < \lambda < \infty$. In this section Greek letters (excluding $\lambda$) denote coordinates on $VM$  while Latin letters range 0 to $m-1$ and denote coordinates on $M$.

The Thomas-Whitehead connection $\tilde{\Gamma}^a_{ \ \beta \gamma}$ lives on $VM$, and is both projectively  invariant and houses $\Pi$ as a component. The TW connection on $VM$ can be decomposed as
\begin{equation}
    \tilde{\Gamma}^a_{ \ \beta \gamma} = \begin{cases}
        \tilde{\Gamma}^a_{\ bc} = \tensor{\Pi}{^a_{bc}} \\
        \tensor{\tilde{\Gamma}}{^{\lambda}_{bc}} = \lambda \mathcal{D}_{bc} \\ 
        \tensor{\tilde{\Gamma}}{^a_{b \lambda}} = \tensor{\tilde{\Gamma}}{^a_{ \lambda b}} = \frac{1}{\lambda} \tensor{\delta}{^a_b} \\ 
        \tensor{\tilde{\Gamma}}{^{\lambda}_{b \lambda}} = \tensor{\tilde{\Gamma}}{^{\lambda}_{ \lambda b}} = \tensor{\tilde{\Gamma}}{^{\lambda}_{ \lambda \lambda}} = 0 \\ 
    \end{cases}\, ,
\end{equation}
where $\mathcal{D}_{bc}$ is a (non-tensorial) rank 2 object on $M$.  In general this need not be related to the Ricci tensor and when  related to the Virasoro algebra  in the literature, it is known as the diffeomorphism (Diff) Field \cite{Rodgers:1994ck}. Demonstrating this component of the TW connection appears in the geometric action of the Diffeomorphism group of $S^1$ was the essence of \cite{Brensinger:2017gtb}. For $\tilde{\Gamma}$ to be a connection on $VM$ it must transform as 
\begin{align}
\label{eq: TW connection transformation}
\tensor{\tilde{\Gamma}}{^\alpha_{\beta\gamma}} \to \frac{\partial y^\alpha}{\partial x^\delta} \frac{\partial x^\epsilon}{\partial y^\beta} \frac{\partial x^\eta}{\partial y^\gamma} \tensor{\tilde{\Gamma}}{^\delta_{\epsilon\eta}} + \frac{\partial y^\alpha}{\partial x^\delta} \frac{\partial^2 x^\delta}{\partial y^\gamma\partial y^\beta},    
\end{align}
under coordinate transformations on $VM$. For this to happen the transformation law for the diffeomorphism field must be \cite{Thomas:1925b,Thomas:1925a,Brensinger:2020gcv} 
\begin{align}
\hat{\mathcal{D}}_{bc}=\Bigg(\mathcal{D}_{ef}-\frac{\partial}{\partial x^e}j_f+\tensor{\Pi}{^d_{ef}}j_d-j_ej_f\Bigg)\frac{\partial x^e}{\partial y^b}\frac{\partial x^f}{\partial y^c},    
\end{align}
where $j_a = \partial_a \log{J^{-\frac{1}{m+1}}}$. We emphasize that $\mathcal{D}_{bc}$ lives exclusively on $M$ and is required to ensure that $\tilde{\Gamma}$ transform as a connection over $VM$.

\section{A Superquick Review of Supermanifolds}

This section is based primarily on DeWitt's book \cite{DeWitt:2012mdz} and Rogers' book \cite{Rogers:2007zza}, as well as the  \cite{Rogers:1979vp}, where a Rogers supermanifold is defined.

\subsection{Construction of Supernumbers}

Let $\theta^a$ denote the generators of an algebra subject to the relation
\begin{align}
\theta^a\theta^b&=-\theta^b\theta^a,    
\end{align}
where $a,b=1,...,N.$ This algebra is the $N$-dimensional Grassmann algebra $\Lambda_N$. As a vector space over $\mathbb{C}$, $\Lambda_N(\mathbb{C})$ is $2^N$-dimensional with a basis given by
\begin{align}
&1, \theta^a, \theta^a\theta^b,\theta^a\theta^b\theta^c,...,\theta^1\theta^2...\theta^{N-1}\theta^N.
\end{align}
Throughout we restrict our attention to the field of complex numbers and denote $\Lambda_N(\mathbb{C})$ by $\Lambda_N$. Taking $N\rightarrow \infty$ we obtain the infinite-dimensional Grassmann algebra $\Lambda_\infty$. 

A supernumber $z\in\Lambda_\infty$ is a sum 
\begin{align}
z & = \sum_{n=0}^\infty\frac{1}{n!}c_{a_1...a_n}\theta^{a_n}...\theta^{a_1},
\end{align}
where $c_{a_1...a_n}\in \mathbb{C}$. A useful decomposition of supernumbers is given by the splitting 
\begin{align}
z&=u+v,\\
u&= \sum_{n=0}^\infty\frac{1}{(2n)!}c_{a_1...a_{2n}}\theta^{a_{2n}}...\theta^{a_1},\\
v&=\sum_{n=0}^\infty\frac{1}{(2n+1)!}c_{a_1...a_{2n+1}}\theta^{a_{2n+1}}...\theta^{a_1},
\end{align}
where $u$ and $v$ are the even and odd parts of $z$ and are called $c$-numbers and $a$-numbers, respectively. The set of $c$-numbers and $a$-numbers are denoted by $\mathbb{C}_c$ and $\mathbb{C}_a$, respectively, and are $2^{N-1}$-dimensional vector spaces over $\mathbb{C}$. $\mathbb{C}_c$ is a subalgebra of $\Lambda_\infty$, while $\mathbb{C}_a$ is not as it is not closed under multiplication. 

The real counterparts of $\mathbb{C}_c$ and $\mathbb{C}_a$ are denoted by $\mathbb{R}_c$ and $\mathbb{R}_a$, respectively and are introduced by defining complex conjugation in the following fashion:
\begin{align}
(z_1+z_2)^*&=z_1^*+z_2^*,\\
(z_1z_2)^*&=z_2^*z_1^*,\\
\theta^{a*}&=\theta^a,
\end{align}
where $z_i\in\Lambda_{\infty}$ and $\theta^a$ are generators of $\Lambda_{\infty}$. A supernumber $z$ is real if $z^*=z$ and imaginary if $z^*=-z$.

\subsection{Supervectors and Supermatrices}

Here we present the most salient features of supervectors and supermatrices we will encounter. The usual properties of vector spaces persist in the graded setting, except scalar multiplication is now distinct when acting on the left and right. This can be appreciated via the  decomposition
\begin{align}
X&=U+V,\\
\alpha U&=U\alpha,\\
\alpha V&=-V\alpha,
\end{align}
where $\alpha \in \mathbb{C}_a$ and $U$, $V$ are called the even and odd parts of $X$, respectively. If the odd (even) part of a supervector is not present the supervector is said to be $c$-type ($a$-type). Supernumbers and supervectors of a definite type are called pure. When both the supernumber and the supervector are pure this becomes $ \alpha X =(-1)^{\alpha X} X \alpha $, with the power of $-1$ reflecting the parity. Let $\{_i\bf e\}$ denote a discrete set of basis supervectors and  their duals $\{\bf e^i\}$ such  that $_i{\bf e}\cdot {\bf e}^j = \ _i\tensor{\delta}{^j}$.  Then a supervector may be denoted as  $X= X^i \, _i{\bf e} $. Here we also introduce left and right derivations,  $_ae=\frac{\cev{\partial}}{\partial x^a}$ and $e_a=\frac{\vec{\partial}}{\partial x^a}$. 
Therefore $X^i\, _i{\bf e}= X^i \frac{\vec{\partial}}{\partial x^i}= U - V$, while $X^i\, {\bf e}_i= X^i \frac{\cev{\partial}}{\partial x^i}= U + V.$  If the object of interest is $c$-type ($a$-type), then its associated symbol equals 0 (1). For pure supervectors, one convenience is to write  $^i X \equiv\ (-1)^{Xi} \ X^i$, so in particular for $c$-type supervectors $^i X = X^i$. The advantage of writing $X = X^i\, _i{\bf e}$ lies in the minimization of parity factors.

Let $_j K^i$ be a matrix of supernumbers. The components of a supervector transform by $\bar{X}^i = X^j {_jK^i}$, as summation is carried out only with adjacent indices. For $K$ to preserve the parity of $X$ it must be a block matrix of the form 
\begin{align}
K=\begin{pmatrix} A& C\\ D& B\end{pmatrix},    
\end{align}
where $A$ and $B$ are comprised of $c$-numbers and $C$ and $D$ are comprised of $a$-numbers. Such matrices are $c$-type precisely because they preserves the type of the supervector they act on and thus introduce no parity of their own. For $c$-type supermatrices we can form the product, supertranspose, and supertrace, respectively, as 
\begin{align}
_i(KL)^j  ={_iK^k}{_kL^j}&\,\,\,\,\text{[product rule]} \nonumber \\ 
^i{K^\sim}_j  = (-1)^{j(i+j)}{_jK^i}=(-1)^{j(i+1)}{_jK^i} &\,\,\,\,\text{[supertranspose] } \nonumber \\
\text{str}(K) = \tensor{K}{_i^i} = (-1)^i \, {_iK^i} &\,\,\,\,\text{[supertrace]},
\end{align}
where $\sim$ denotes the supertranspose operation. 
The superdeterminant is defined for any supermatrix and enjoys the usual  multiplication laws. If $K$ has the index structure $_i K^j$ or $^i K_j$ then
\begin{align}
\text{sdet}(K^\sim)&=\text{sdet}(K),
\end{align} 
while if $K$ has index structure $_i K_j$ or $^i K^j$ then
\begin{align} 
\text{sdet}(M^\sim)&=(-1)^n\text{sdet}(M). 
\end{align}
Let $A$, $B$, $C$, and $D$ be $m\times m$, $n\times n$, $m\times n$, $n\times m$, $m\times n$, and $n\times m$ matrices with entries comprised of $c$, $c$, $a$, and $a$ numbers, respectively. Then 
\begin{align}
& \text{sdet}\begin{pmatrix} A& C\\ D& B\end{pmatrix}=\text{det}(A-CB^{-1}D)(\text{det}\,B)^{-1}, 
\end{align}
where $\text{sdet}$ is defined only if $B$ is invertible. The superdeterminant of a $c$-type matrix is a real $c$-number. See \cite{DeWitt:2012mdz} for a treatment of the superdeterminant for $a$-type matrices.

\section{The super Thomas-Whitehead connection}

\subsection{Super Coordinate and Projective Transformations}

In the graded setting we consider the coordinate transformation $x^a\rightarrow y^a$ with the super Jacobian and its inverse 
\begin{align}
^a J_b & = (-1)^{b(a+b)} {_{b,}y^a}=(-1)^{b(a+b)} \frac{\vec{\partial}}{\partial x^b}y^a, \\ 
^a{(J^{-1})}_b & = x^a \frac{\cev{\partial}} {\partial y^b} = (-1)^{b(a+b)} {_{b,}x^a}=(-1)^{b(a+b)} \frac{\vec{\partial}}{\partial y^b}x^a, \nonumber 
\end{align}

which satisfies analogues of the usual identities 
\begin{align} 
^a{(J^{-1}})_b 
\ {^bJ_{c}} & = {^a\delta_c} = x^a \frac{\cev{\partial}}{\partial x^c}, \\
{_cJ^b} \ {_b(J^{-1})^a} & = {_c\delta^a} = \frac{\vec{\partial}}{\partial x^c} x^a,\\
\text{and} \,\,\,\, \tensor{x}{^a_{,bc}} & = {^a{(J^{-1})}_b} \frac{\cev{\partial}}{\partial x^d}{^d{(J^{-1})}_c}.
\end{align}

There are two graded analogs of the Jacobi formula, one for left and one for right derivatives, respectively  \cite{DeWitt:2012mdz}: 
\begin{align}
\frac{\vec{\partial}}{\partial x^a} \text{ln}(\text{sdet}(M)) & = \text{str}((M{_ae})M^{-1}) = \text{str}(M^{-1}(M{_ae})), \nonumber \\ 
\text{ln}(\text{sdet}(M))\frac{\cev{\partial}}{\partial x^a} & = (-1)^i \, {^iM_j}\frac{\cev{\partial}}{\partial x^a}{^j(M^{-1})_i} =  (-1)^i \, {^i(M^{-1})_j}{^jM_i}\frac{\cev{\partial}}{\partial x^a}. 
\end{align}

The relationship between left and right derivatives can be used to show that the supertrace is invariant under cyclic permutations. A super projective transformation is analogous to the ungraded case and is given by the relation 
\begin{align}
\tensor{\hat{\Gamma}}{^a_{bc}}=\tensor{\Gamma}{^a_{bc}}+{^a\delta_b}v_c+{^a\delta_c}(-1)^{bc}v_b,    
\end{align}
where $v_a$ are the components of a $c$-type 1-form.

\subsection{The Super Fundamental Projective Invariant}

The fundamental projective invariant, $\Pi$, can be promoted to a graded setting by replacing the trace with the supertrace and adding necessary parity factors \cite{JGeorge}. The resulting geometrical object is the super fundamental projective invariant
\begin{align}
\tensor{\Pi}{^a_{bc}} \equiv \tensor{\Gamma}{^a_{bc}} - D \big( {^a\delta_b}(-1)^e \tensor{\Gamma}{^e_{ec}} + {^a\delta_c}(-1)^{e+bc} \tensor{\Gamma}{^e_{eb}}\big),
\end{align}
where we set $D\equiv(m-n+1)^{-1}$ for future convenience. The coordinate transformation law of the super fundamental projective invariant is:
\begin{align}
\tensor{\bar{\Pi}}{^a_{bc}}&={^aJ_d}\Big((-1)^{g(f+b)}\tensor{\Pi}{^d_{fg}}{^f(J^{-1})_b}{^g(J^{-1})_c}+\tensor{x}{^d_{,bc}}\Big)\\
& + D \Big({^a\delta_b}\text{ln}(J)\frac{\cev{\partial}}{\partial x^f}{^f(J^{-1})_c}+{^a\delta_c}(-1)^{bc}\text{ln}(J)\frac{\cev{\partial}}{\partial x^f}{^f(J^{-1})_b}\Big).\nonumber
\end{align}

\subsection{The TW Connection}
The coefficients  of $\tensor{\tilde{\Gamma}}{^\alpha_{\beta \gamma}}$ in the graded setting are: 
\begin{align}
\tensor{\tilde{\Gamma}}{^a_{bc}}&=\tensor{\Pi}{^a_{bc}}=(-1)^{bc}\tensor{\Pi}{^a_{cb}}, \\
\tensor{\tilde{\Gamma}}{^{\lambda}_{bc}} &= \lambda \mathcal{D}_{bc}=(-1)^{bc} \lambda \mathcal{D}_{cb},\\
\tensor{\tilde{\Gamma}}{^a_{b \lambda}}&=\tensor{\tilde{\Gamma}}{^a_{ \lambda b}}=\frac{1}{\lambda}\tensor{\delta}{^a_b},
\end{align}
where any components not listed vanish. The measure transforms as   
\begin{align}
d^{m,n}x\rightarrow d^{m,n}\bar{x}=Jd^{m,n}x,
\end{align}
under a coordinate transformation $x^i\rightarrow \bar{x}^i(x),$ where $J\equiv\text{sdet}(\tensor{\bar{x}}{^i_{,j}})$ is the super Jacobian or Berezinian, and where the number of $a$-type coordinates must be even for the metric to be nonsingular. Latin indices range over the supermanifold coordinates while Greek indices (except $\lambda$) range over all coordinates, $\lambda, x^a$. 

As before, we check the connection coefficients recover the transformation law for $\Pi$. The graded extensions of the identities from before are: 
\begin{align}
&j_a\equiv \text{log}(J^{-D})\frac{\cev{\partial}}{\partial x^a} = \text{log}(J^{-D})_{,a} = (\lambda J^{-D}) \frac{\cev{\partial}}{\partial x^a} \frac{1}{(\lambda J^{-D})},\\
&\lambda \frac{\cev{\partial}}{\partial y^a} \frac{1}{\lambda}  = \text{log}(J^D) \frac{\cev{\partial}}{\partial x^g} {^g(J^{-1})_a} = - j_g{^g(J^{-1})_a},\\
&{^{\lambda}(J^{-1})_a}  = - \lambda j_g \ {^g(J^{-1})_a}, \,\,\,\,\text{and}\\
&{^{\lambda} J_a} = \lambda J^{-D} j_a. \end{align}

With this, the transformation law for the super TW connection is 
\begin{align}
\tensor{\tilde{\Gamma}}{^\alpha_{\beta\gamma}} \to {^\alpha J_\delta} \tensor{x}{^\delta_{,\beta\gamma}} + (-1)^{\eta(\epsilon + \beta)} \ {^\alpha J_\delta} \tensor{\tilde{\Gamma}} {^\delta_{\epsilon\eta}} {^\epsilon(J^{-1})_\beta}{^\eta(J^{-1})_\gamma}.
\end{align}
The coordinate transformation law for the super Diffeomorphism field is 
\begin{align}
\mathcal{D}_{bc} \to (-1)^{f(b+e)} \Big( \mathcal{D}_{ef} - j_{e,f} - j_e j_f + j_d \tensor{\Pi} {^d_{ef}} \Big) {^e(J^{-1})_b} {^f(J^{-1})_c}. 
\end{align}
The parity of $\mathcal{D}_{ab}$ is $(-1)^{a+b}$. Under an infinitesimal coordinate transformation $x^a \to x^a - \delta \epsilon^a, $ 
\begin{equation}
\label{eq: Graded Diff transformation}
\mathcal{D}_{bc}(x) \to \mathcal{D}_{bc} (x) + \delta \bigg( \mathcal{D}_{bc,i} \epsilon^i + \mathcal{D}_{bf} (x) \tensor{\epsilon}{^f_{,c}} + (-1)^{c(b+e)} \mathcal{D}_{ec} (x) \tensor{\epsilon}{^e_{,b}}-D(-1)^i \tensor{\epsilon}{^i_{,ibc}} + D(-1)^i\tensor{\epsilon}{^i_{,id}} \tensor{\Pi}{^d_{bc}} (x) \bigg), \nonumber
\end{equation}
where the coefficient of $\delta$ is the super Lie derivative with respect to the vector field $\epsilon$. Setting the fermionic dimension to zero and the bosonic dimension to one and noting that the super Diffeomorphism has a single component,  we recover  the following reduction:
\begin{align}
\mathcal{D} \to \mathcal{D} + \delta \bigg( \mathcal{D}' \epsilon + 2 \mathcal{D} \epsilon' - \frac{1}{2} \epsilon''' \bigg). 
\end{align}
After a redefinition of $\mathcal{D}$, we recover the coordinate transformation  on a coadjoint Virasoro element \cite{Brensinger:2020gcv,Brensinger:2017gtb}. 
It is sometimes convenient to express $\Pi$ in terms of a member in the equivalence class  of  connections and subtracting out the trace of this member which we denote by $\alpha$. In general, our torsionless affine connection $\Gamma$ will not be compatible with the metric.
 In terms of this member, we have:\begin{align}
\alpha_a&\equiv -D(-1)^e\tensor{\Gamma}{^e_{ea}},\\ \tensor{\Pi}{^a_{bc}}&=\tensor{\Gamma}{^a_{bc}}+\tensor{\delta}{^a_b}\alpha_c+(-1)^{bc}\tensor{\delta}{^a_c}\alpha_b.
\end{align}
We may then express the projective Ricci symbol in terms of this connection and and its trace and write,
\begin{align}
\mathcal{R}_{bd}&=(-1)^{c(b+c+d)}\tensor{\Gamma}{^c_{bd,c}}-(-1)^{d(b+c)+c}\tensor{\Gamma}{^c_{df}}\tensor{\Gamma}{^f_{cb}}\\
&+(-1)^{db}(\alpha_{d,b}-\alpha_f\tensor{\Gamma}{^f_{db}}) + \alpha_{b,d}-\alpha_f\tensor{\Gamma}{^f_{bd}} + (m-n-1) \alpha_b \alpha_d.\nonumber
\end{align}
If $\Gamma$ is Levi-Civita then $\alpha_a=-D^{-1}\text{log}(g^{1/2})_{,a}$ and
\begin{align}
(-1)^{db}\alpha_{d,b} = \alpha_{b,d}.    
\end{align}
The super projective Ricci symbol transforms as the super Diffeomorphism field up to a constant. This provides us with an alternative way to deduce the parity of $\mathcal{D}$. Furthermore, we may define a tensor on $M$ defined as
\begin{align}
\mathcal{P}_{ab} = \tensor{\mathcal{D}}{_{ab}} - \alpha_{a,b} + \alpha_f \tensor{\Gamma}{^f_{ab}} + \alpha_a \alpha_b.
\end{align}

$\mathcal{P}_{ab}$ is a rank 2 $c$-type tensor with parity $(-1)^{a+b}$, and is known to differential geometers in the ungraded setting as the Projective Schouten tensor. If our connection is Levi-Civita then $\mathcal{P}$ is supersymmetric, i.e. $\mathcal{P}_{ab} = (-1)^{ab}\mathcal{P}_{ba}.$ By inserting a parity factor and contracting over the 1\textsuperscript{st} and 3\textsuperscript{rd} indices of the super projective Riemann curvature symbol, we obtain the super projective Ricci symbol from before except that $\mathcal{P}$ and $\mathcal{D}$ are now present.
\begin{align}
\mathcal{R}_{bd}&=(-1)^{c(b+c)}\tensor{\mathcal{R}}{^c_{bcd}}\nonumber\\
&=R_{bd}+(m-n-1)\big(\mathcal{P}_{bd}-\mathcal{D}_{bd}\big)+(-1)^{bd}\alpha_{d,b}-\alpha_{b,d}.
\end{align}
Rearranging the above, we have the following form for the super Diffeomorphism field:
\begin{align}
\mathcal{D}_{bd}&=-\frac{1}{m-n-1}\mathcal{R}_{bd}+\bigg(\frac{1}{m-n-1}\big(R_{bd}+(-1)^{bd}\alpha_{d,b}-\alpha_{b,d}\big)+\mathcal{P}_{bd}\bigg).
\end{align}
One can always shift the super Diffeomorphism field by any symmetric rank-2 tensor on $M$.

\section{The super Thomas-Whitehead curvature tensor}
It is natural to form a curvature tensor from the super TW connection, which we call the super TW curvature tensor or the super projective Riemann curvature tensor.
\begin{align}
\tensor{\mathcal{K}}{^\alpha_{\beta \gamma \delta}}&=-\tensor{\tilde{\Gamma}}{^\alpha_{\beta\gamma,\delta}}+(-1)^{\gamma \delta}\tensor{\tilde{\Gamma}}{^\alpha_{\beta \delta,\gamma}}-(-1)^{\delta(\epsilon+\beta+\gamma)}\tensor{\tilde{\Gamma}}{^\alpha_{\epsilon\delta}}\tensor{\tilde{\Gamma}}{^\epsilon_{\beta \gamma}}+(-1)^{\gamma(\epsilon+\beta)}\tensor{\tilde{\Gamma}}{^\alpha_{\epsilon\gamma}}\tensor{\tilde{\Gamma}}{^\epsilon_{\beta \delta}}.
\end{align}
Before discussing how $\tensor{K}{^\alpha_{\beta \gamma \delta}}$ transforms, we recall how a tensor of rank $(1,3)$ transforms under a change of basis.
\begin{align}
\tensor{\bar{T}}{^{a_1}_{a_2a_3a_4}} & = (-1)^{\Delta_{4}(a+b,b)} \tensor{T}{^{b_1}_{b_2b_3b_4}}{_{b_1}(L^{-1\sim})^{a_1}}{^{b_2}L_{a_2}}{^{b_3}L_{a_3}}{^{b_4}L_{a_4}},
\end{align}
where $\Delta$ for a tensor of rank $(r,s)$ is defined as ($q=r+s$) 
\begin{align}
\Delta_q(a,b)\equiv \sum_{\substack{t,u=1\\ t<u}}^na_tb_u.  
\end{align}
Hence, $\tensor{\mathcal{K}}{^\alpha_{\beta \gamma \delta}}$ transforms as 
\begin{align}
\tensor{\mathcal{K}}{^{\alpha_1}_{\alpha_2\alpha_3\alpha_4}} \to (-1)^{\Delta_4(\alpha+\beta,\beta)} \tensor{\mathcal{K}}{^{\beta_1}_{\beta_2\beta_3\beta_4}}{_{\beta_1} (L^{-1\sim})^{\alpha_1}}{^{\beta_2}L_{\alpha_2}}{^{\beta_3}L_{\alpha_3}}{^{\beta_4}L_{\alpha_4}}.  
\end{align}
The non-vanishing components of $\mathcal{K}^{\alpha}_{\ \beta \gamma \delta}$ are
\begin{align}
\tensor{\mathcal{K}}{^{\lambda}_{abc}} & = \lambda \big((-1)^{bc} \mathcal{D}_{ac,b} - \mathcal{D}_{ab,c}+(-1)^{b(a+d)} \mathcal{D}_{db} \tensor{\Pi}{^d_{ac}} - (-1)^{c(a+b+d)} \mathcal{D}_{dc} \tensor{\Pi}{^d_{ab}} \big),\\
\tensor{\mathcal{K}}{^a_{bcd}} & = \tensor{\mathcal{R}}{^a_{bcd}} + (-1)^{bc} \tensor{\delta}{^a_c}\mathcal{D}_{bd}-(-1)^{d(b+c)}\tensor{\delta}{^a_d}\mathcal{D}_{bc} \nonumber \\ 
&=\tensor{R}{^a_{bcd}}-\tensor{\delta}{^a_b}\big(
(-1)^{cd}\mathcal{P}_{dc}-\mathcal{P}_{cd}\big) + (-1)^{bc}\tensor{\delta}{^a_c}\mathcal{P}_{bd}-(-1)^{d(b+c)}\tensor{\delta}{^a_d}\mathcal{P}_{bc}.\nonumber 
\end{align}
where  $\tensor{K}{^a_{bcd}}$ is called the super projective Riemann curvature tensor on M. For convenience we also introduce the tensor 
\begin{align}
    \Breve{\mathcal{K}}_{abc} & = \frac{1}{\lambda} \tensor{\mathcal{K}}{^{\lambda}_{abc}}.
\end{align}
Under a super projective transformation $\mathcal{P}$ transforms as \begin{align}
\mathcal{P}_{ab} & \rightarrow \mathcal{P}_{ab} + v_{a;b} - v_a v_b.
\end{align}
Rewriting the other components of the super Thomas-Whitehead curvature tensor, we have 
\begin{align}
\tensor{\Breve{\mathcal{K}}}{_{abc}} & = (-1)^{cb} \mathcal{P}_{ac,b} - \mathcal{P}_{ab,c} + (-1)^{cb} \alpha_a \mathcal{P}_{cb}-\alpha_a \mathcal{P}_{bc} + (-1)^{b(a+d)} \mathcal{P}_{db}\tensor{\Gamma}{^d_{ac}} \\
& - (-1)^{c(a+b+d)}\mathcal{P}_{dc}\tensor{\Gamma}{^d_{ab}} +\mathcal{P}_{ab}\alpha_c-(-1)^{bc}\mathcal{P}_{ac}\alpha_b-\alpha_f\tensor{R}{^f_{abc}}.\nonumber
\end{align}
Similarly, the super TW Ricci tensor is formed by contracting the 1\textsuperscript{st} and 3\textsuperscript{rd} indices of the super TW curvature tensor 
\begin{align}
\mathcal{K}_{bd} & = R_{bd}+(m-n)\mathcal{P}_{bd}-(-1)^{db}\mathcal{P}_{db}.
\end{align}
We can write the super TW Ricci tensor in terms of the connection by taking the trace 
\begin{align}
R_{jl} & = (-1)^{k(j+1)} \tensor{R}{^k_{jkl}} \nonumber 
\end{align}
and the super TW Ricci scalar as 
\begin{align}
\mathcal{K} = \mathcal{K}_{ab} g^{ba} = R+(m-n-1) \mathcal{P}.
\end{align}

\section{The super Thomas-Whitehead action}

The TW action \cite{Brensinger:2017gtb,Brensinger:2019mnx,Brensinger:2020gcv} is constructed from the sum of the projective Einstein-Hilbert and projective Gauss-Bonnet Lagrangians. In this section we construct the analogous super TW action  as the sum of the super projective Einstein-Hilbert Lagrangian, $\mathcal{L}_{SPEH}$, and the super projective Gauss-Bonnet Lagrangian, $\mathcal{L}_{SPGB}$. In the TW action, $g_{ab}$, $\Pi^{a}_{\,\,b c}$, and $\mathcal{D}_{ab}$ are all independent field degrees of freedom, where the metric serves to build coordinate invariant objects. 

In the super TW action, $\mathcal{L}_{SPEH}$ generates the super Einstein-Hilbert term and couples the  metric on $M$ and the super Diffeomorphism field. The tensor $\breve{\mathcal{K}}_{abc}$ contains   $\mathcal{D}$, $\Pi$, and derivatives on $\mathcal{D}$. The square of $\breve{\mathcal{K}}_{abc}$ sources dynamics of $\mathcal{D}$ that arises in the projective Gauss-Bonnet action. In the limit where $\Pi$ is compatible with the metric and the diff field vanishes the TW action collapses to the Einstein-Hilbert action. This follows from the fact that the Gauss-Bonnet action is a topological invariant in four dimensions. Furthermore, it is known that in any dimension, $\mathcal{L}_{SPGB}$ has only second-order derivatives of the metric \cite{Lanczos:1938sf,Lovelock:1971yv}, which keeps the field equations from developing higher time derivatives.  Finally, we emphasize that the TW Lagrangian over $VM$ is invariant under both super coordinate and super projective transformations.

The first task is to endow our supermanifold $M$ with a metric $g$, and then promote $g$ to a metric $G$ on $VM$. Let's recall the structure of a $c$-type matrix $k$ that acts on an $(m,n)$-dimensional supervector space:
\begin{align}
k=\begin{pmatrix}A_{m\times m}& C_{m\times n}\\D_{n\times m}&B_{n\times n}\end{pmatrix},
\end{align}
where $A$, $B$, $C$, and $D$ are of type $c$, $c$, $a$, and $a$, respectively. For the Thomas cone we need to increase the dimension of the bosonic sector by one and arrange the decomposition of the matrix in order to showcase both the pure and mixed subsectors 
\begin{align}
K=\begin{pmatrix}A^1_{1\times 1}&A^2_{1\times m}&C^1_{1\times n}\\A^3_{m\times 1}&A^4_{m\times m}&C^2_{m\times n}\\D^1_{n\times 1}&D^2_{n\times m}&B_{n\times n}\end{pmatrix}.    
\end{align}
Even though we have changed the bosonic dimension and offered a refined decomposition, the components of $A^i$, $B^j$, $C^k$, and $D^l$ are still of type $c$, $c$, $a$, and $a$, respectively. We promote the metric over $VM$ \cite{Brensinger:2020gcv,Brensinger:2017gtb} to the graded setting
\begin{align}
{_\mu G_\nu} = 
    \begin{pmatrix} 
    - \frac{\lambda^2_0}{\lambda^2} & -  \frac{\lambda^2_0}{\lambda} g_b & - \frac{\lambda^2_0}{\lambda} g_B \\ 
    -\frac{\lambda^2_0}{\lambda}{_ag} & {_ag_b}-\lambda^2_0 \, {_ag}g_b&{_ag_B} - \lambda^2_0 \, {_ag}g_B \\ 
    -\frac{\lambda^2_0}{\lambda}{_Ag}&{_Ag_b}-\lambda^2_0{_Ag}g_b&{_Ag_B}-\lambda^2_0{_Ag}g_B
    \end{pmatrix} =
    \begin{pmatrix}
    {_{\lambda} G_{\lambda}}&{_{\lambda} G_b}&{_{\lambda} G_B} \\ 
    { _aG_{\lambda}}&{_aG_b}&{_aG_B} \\ 
    {_AG_{\lambda}} & {_AG_b} & {_AG_B} 
    \end{pmatrix},    
\end{align}
where $a$ and $b$ range over the even coordinates (except $\lambda$), $A$ and $B$ range over the odd coordinates, $\mu$ and $\nu$ range over all coordinates, and $g_a\equiv -D^{-1}\text{log}(g^{1/2})_{,a}=(-1)^a{_a g}.$ If we choose the Levi-Civita connection, then $g_a=\alpha_a$. $\lambda_0$ is introduced in order to render the components of the metric dimensionless and $g_a$ has units of inverse length. 

Considering the case $ M = \mathbb{R}^m_c \times \mathbb{R}^n_a$, the metric simplifies considerably as $g_{aA} = g_{Aa} = 0$ and $g_a=g_A=0.$ The canonical form of the metric on $\mathbb{R}^2_c \times \mathbb{R}^2_a$ is \cite{DeWitt:2012mdz} 
\begin{align}
\eta = 
    \begin{pmatrix}
    -1&0&0&0 \\ 
    0&1&0&0\\ 
    0&0&0&i\\ 
    0&0&-i&0 
    \end{pmatrix} = \begin{pmatrix} 
    {_a\eta_b} & 0 \\ 
    0&{_A\eta_B}
    \end{pmatrix},    
\end{align}

implying the metric on the volume bundle of $ \mathbb{R}^m_c \times \mathbb{R}^n_a$ is 
\begin{align}
{_\mu G_\nu}= 
    \begin{pmatrix}
    - \frac{\lambda^2_0}{\lambda^2} & 0 & 0 \\ 
    0 & {_a \eta_b} & 0 \\ 
    0 & 0 & {_A \eta_B} 
    \end{pmatrix}. 
\end{align}

We have the relationship between the metric on $M$ and on $VM$
\begin{align}
\text{sdet} \big({_\mu G_\nu}\big) = - \frac{\lambda^2_0}{\lambda^2}  \text{sdet} ({_{\dot{\mu}} g_{\dot{\nu}}}),
\end{align}
where $\dot{\mu}$, $\dot{\nu}$ range over all coordinates except $\lambda$. $G$ satisfies $G^{\mu\nu}=(-1)^{\mu\nu}G^{\nu\mu}$ and transforms on $VM$ as
\begin{align}
{G}_{\mu\nu} \to (-1)^{\sigma(\rho+\mu)} G_{\rho \sigma} \ {^\rho(L^{-1})_\mu}{^\sigma(L^{-1})_\nu},    
\end{align}
where the metric with both of its indices to the bottom right takes the form 
\begin{align}
G_{\mu\nu} = 
    \begin{pmatrix}
    - \frac{\lambda^2_0}{\lambda^2} & 0 & 0 \\
    0 & g_{ab} & 0 \\
    0 & 0 & g_{AB} \end{pmatrix} = \begin{pmatrix}
    - \frac{\lambda^2_0}{\lambda^2} & 0 \\
    0 & g_{\dot{\mu}\dot{\nu} }\end{pmatrix} (-1)^\mu{_\mu G_\nu}.    
\end{align}
This metric is symmetric and invariant under super projective transformations by construction. The inverse is 
\begin{align}
G^{\mu\nu} = 
    \begin{pmatrix}
    - \frac{\lambda^2}{\lambda^2_0} & 0 & 0\\ 
    0 & g^{ab} & 0 \\ 
    0 & 0 & g^{AB} 
    \end{pmatrix} = \begin{pmatrix} 
    - \frac{\lambda^2}{\lambda^2_0} & 0 \\ 
    0 & g^{\dot{\mu}\dot{\nu}} 
    \end{pmatrix}
={^\mu G^\nu}.    
\end{align}

Now that we have a metric on $VM$ we are ready to construct an action. We revert back to our old convention where Greek and Latin indices range over the coordinates of $VM$ and $M$, respectively. This change causes the metric on $VM$ to take the shape
\begin{align}
{_\mu G_\nu} = 
    \begin{pmatrix}{_{\lambda} G_{\lambda}} & {_{\lambda} G_b} \\ 
    {_aG_{\lambda}} & {_aG_b} \end{pmatrix} = \begin{pmatrix} - \lambda^{-2}&-\lambda^{-1}g_b\\-\lambda^{-1}{_ag}&{_ag_b}-{_ag}g_b\end{pmatrix},    
\end{align}
with inverse given by 
\begin{align}
{^\nu G^\rho}=\begin{pmatrix}\lambda^2(g_m{^mg^n}{_ng}-1)&-\lambda g_m{^mg^c}\\-\lambda{^bg^m}{_mg}&{^bg^c}\end{pmatrix},    
\end{align}

Our next task is to construct the square of the super projective Riemann curvature tensor, which has 12 parity terms. Shifting the metrics to the left will result in many more parity terms:
\begin{align}
\tensor{\mathcal{K}}{^\alpha_{\beta\gamma\delta}} \tensor{\mathcal{K}}{_\alpha^{\beta\gamma\delta}} = (-1)^{\rho( \mu + \nu + \sigma + \alpha + \beta) + \sigma( \mu + \nu + \alpha + \beta + \gamma) + \nu( \mu + \alpha)} \tensor{\mathcal{K}}{^\alpha_{\beta\gamma\delta}} \tensor{\mathcal{K}}{^\mu_{\nu\rho\sigma}}{_\mu G_\alpha} G^{\nu \beta} G^{\rho \gamma} G^{\sigma\delta} 
\end{align}
The next summand we need is the  square of the super projective Ricci tensor, given by
\begin{align}
\mathcal{K}_{\beta\delta} \mathcal{K}^{\beta\delta} & = (-1)^{(\nu+\beta)\delta} \mathcal{K}_{\beta\delta} \mathcal{K}_{\nu\sigma} G^{\sigma\delta} G^{\nu\beta},
\end{align}
which can be written in terms of its ancestor, the super projective Riemann curvature tensor, as 
\begin{align}
\mathcal{K}_{\beta\delta} \mathcal{K}^{\beta\delta} & = (-1)^{\gamma(\nu+\sigma+\beta+\gamma)+\alpha(\nu+\sigma)+\rho(\nu+\rho)+\sigma(\nu+\beta)}\tensor{\mathcal{K}}{^\alpha_{\beta\gamma\delta}}\tensor{\mathcal{K}}{^\mu_{\nu\rho\sigma}}\tensor{\delta}{^\rho_\mu}\tensor{\delta}{^\gamma_\alpha}G^{\nu\beta}G^{\sigma\delta}.
\end{align}
Finally, we need the square of the super projective Ricci scalar 
\begin{align}
\mathcal{K}^2 &=(-1)^{\gamma(\beta+\gamma)+\rho(\nu+\rho)+(\nu+\sigma)(\alpha+\gamma)}\tensor{\mathcal{K}}{^\alpha_{\beta \gamma \delta}}\tensor{\mathcal{K}}{^\mu_{\nu \rho \sigma}}\tensor{\delta}{^\rho_\mu}\tensor{\delta}{^\gamma_\alpha}G^{\sigma\nu}G^{\delta\beta}.
\end{align}

If we change the order of the tensors in this expression more parity factors will arise, so our convention in expressing candidates for $\mathcal{L}$ will begin with products of $K$, followed by $\delta$ and then $G$. Bringing everything together, we have the super projective Einstein-Hilbert Lagrangian and the super projective Gauss-Bonnet Lagrangian: 
\begin{align}
\mathcal{L}_{SPEH}&=\mathcal{K},\\
\mathcal{L}_{SPGB} & = \tensor{\mathcal{K}}{^\alpha_{\beta\gamma\delta}} \tensor{\mathcal{K}}{_\alpha^{\beta\gamma\delta}} - 4 \mathcal{K}_{\beta\delta} \mathcal{K}^{\beta\delta} + \mathcal{K}^2 \nonumber \\
& = \tensor{\mathcal{K}}{^\alpha_{\beta \gamma \delta}}\tensor{\mathcal{K}}{^\mu_{\nu \rho \sigma}}\tensor{\mathcal{C}}{^{\sigma\delta\rho\gamma\nu\beta}_{\mu\alpha}},
\end{align}
where $\mathcal{C}$ is the super projective Gauss-Bonnet tensor on $VM$ 
\begin{align}
\tensor{\mathcal{C}}{^{\sigma\delta\rho\gamma\nu\beta}_{\mu\alpha}}&=\bigg((-1)^{\delta(\mu+\nu+\rho+\alpha+\beta+\gamma)+\gamma(\mu+\nu+\alpha+\beta)+\beta(\mu+\alpha)+\mu}G^{\sigma\delta}G^{\rho\gamma}G^{\nu\beta}G_{\mu\alpha}\\
&-4(-1)^{\gamma(\nu+\sigma+\beta+\gamma)+\alpha(\nu+\sigma)+\rho(\nu+\rho)+\sigma(\nu+\beta)}\tensor{\delta}{^\rho_\mu}\tensor{\delta}{^\gamma_\alpha}G^{\nu\beta}G^{\sigma\delta}\nonumber\\
&+(-1)^{\gamma(\beta+\gamma)+\rho(\nu+\rho)+(\nu+\sigma)(\alpha+\gamma)}\tensor{\delta}{^\rho_\mu}\tensor{\delta}{^\gamma_\alpha}G^{\sigma\nu}G^{\delta\beta}\bigg).\nonumber    
\end{align}
We define the super Gauss-Bonnet symbol $\mathcal{G}$ on $M$ and the super Gauss-Bonnet tensor $\mathcal{B}$ on $M$ as 
\begin{align}
\tensor{\mathcal{G}}{^{hdgcfb}_{ea}}&\equiv(-1)^{d(e+f+g+a+b+c)+c(e+f+a+b)+b(e+a)+e}g^{hd}g^{gc}g^{fb}(g_{ea}-g_eg_a)\\
&-4(-1)^{c(f+h+b+c)+a(f+h)+g(f+g)+h(f+b)}\tensor{\delta}{^g_e}\tensor{\delta}{^c_a}g^{fb}g^{hd}\nonumber\\
&+(-1)^{c(b+c)+g(f+g)+(f+h)(a+c)}\tensor{\delta}{^g_e}\tensor{\delta}{^c_a}g^{hf}
g^{db},\nonumber\\
\tensor{\mathcal{B}}{^{hdgcfb}_{ea}}&\equiv\tensor{\mathcal{G}}{^{hdgcfb}_{ea}}\\
&=B_1g^{hd}g^{gc}g^{fb}g_{ea}-B_2\tensor{\delta}{^g_e}\tensor{\delta}{^c_a}g^{fb}g^{hd}+B_3\tensor{\delta}{^g_e}\tensor{\delta}{^c_a}g^{hf}
g^{db},
\end{align}
where we have introduced 
\begin{align}
B_1&\equiv(-1)^{d(e+f+g+a+b+c)+c(e+f+a+b)+b(e+a)+e},\\
B_2&\equiv4(-1)^{c(f+h+b+c)+a(f+h)+g(f+g)+h(f+b)},\\
B_3&\equiv(-1)^{c(b+c)+g(f+g)+(f+h)(a+c)}.
\end{align}

We are now ready to expand the super projective Gauss-Bonnet Lagrangian as \begin{align}
\mathcal{L}_{SPCGB} & = \tensor{\mathcal{K}}{^\alpha_{\beta \gamma \delta}} \tensor{\mathcal{K}}{^\mu_{\nu \rho \sigma}} \tensor{\mathcal{C}}{^{\sigma\delta\rho\gamma\nu\beta}_{\mu\alpha}} \\
& = \tensor{\mathcal{K}}{^a_{bcd}} \tensor{\mathcal{K}}{^e_{fgh}} \tensor{\mathcal{G}}{^{hdgcfb}_{ea}} + \tensor{\mathcal{K}}{^{\lambda}_{bcd}} \tensor{\mathcal{K}}{^{\lambda}_{fgh}} \tensor{\mathcal{C}}{^{hdgcfb}_{\lambda \lambda}} \nonumber \\
&+\tensor{\mathcal{K}}{^{\lambda}_{bcd}} \tensor{\mathcal{K}}{^e_{fgh}}\tensor{\mathcal{C}}{^{hdgcfb}_{e \lambda}} + \tensor{\mathcal{K}}{^a_{bcd}} \tensor{\mathcal{K}}{^{\lambda}_{fgh}} \tensor{\mathcal{C}}{^{hdgcfb}_{\lambda a}}. \nonumber
\end{align}
We tackle this Lagrangian one term at a time:
\begin{align}
\mathcal{L}_1 & = \tensor{\mathcal{K}}{^a_{bcd}} \tensor{\mathcal{K}}{^e_{fgh}} \Big(\tensor{\mathcal{B}}{^{hdgcfb}_{ea}} - (-1)^{d(e+f+g+a+b+c) + c(e+f+a+b)+b(e+a)+e} g^{hd} g^{gc} g^{fb} g_e g_a \Big), \\
\mathcal{L}_2 & = - \tensor{\Breve{\mathcal{K}}}{_{bcd}} \tensor{\Breve{\mathcal{K}}}{_{fgh}}(-1)^{d(f+g+b+c)+c(f+b)}g^{hd} g^{gc} g^{fb},\\
\mathcal{L}_3 & = - \tensor{\Breve{\mathcal{K}}}{_{bcd}} \tensor{\mathcal{K}}{^e_{fgh}} (-1)^{d(e+f+g+b+c)+c(e+f+b)+e(b+e)} g^{hd} g^{gc} g^{fb} g_e, \\
\mathcal{L}_4 & = - \tensor{\mathcal{K}}{^a_{bcd}} \tensor{\Breve{\mathcal{K}}}{_{fgh}}(-1)^{d(f+g+a+b+c)+c(f+a+b)+ba} g^{hd} g^{gc} g^{fb} g_a.
\end{align}
Let's  rearrange $\mathcal{L}_3$ to share the same index structure as $\mathcal{L}_4$, and vice versa.
\begin{align}
\mathcal{L}_3 & = - \tensor{\mathcal{K}}{^a_{bcd}} \tensor{\Breve{\mathcal{K}}}{_{fgh}} g^{hd} g^{gc} g^{fb} g_a (-1)^{a+f(c+d)+g(d+f)+h(f+g)} = \tilde{\mathcal{L}}_4,\\
\tilde{\mathcal{L}}_4 + \mathcal{L}_4 & = - \tensor{\mathcal{K}}{^a_{bcd}} \tensor{\Breve{\mathcal{K}}}{_{fgh}} g^{hd} g^{gc} g^{fb} g_a (-1)^{a+f(c+d)+g(d+f)+h(f+g)} \\
& \times \Big( (-1)^{a(a+b+c+d)+b(c+d)+cd+f(g+h)+gh}+1 \Big), \nonumber \\
\mathcal{L}_4 & = \tensor{\mathcal{K}}{^a_{bcd}} \tensor{\Breve{\mathcal{K}}}{_{fgh}} (-1)^{d(f+g+a+b+c)+c(f+a+b)+ba} g^{hd} g^{gc} g^{fb} ( - g_a) = \tilde{\mathcal{L}}_3, \\
\tilde{\mathcal{L}}_3 + \mathcal{L}_3 & = - \tensor{\Breve{\mathcal{K}}}{_{bcd}} \tensor{\mathcal{K}}{^e_{fgh}} g^{hd} g^{gc} g^{fb} g_e (-1)^{d(e+f+g+b+c)+c(e+f+b)+e(b+e)} \\
&\times\Big((-1)^{e(e+f+g+h)+b(c+d)+cd+f(g+h)+gh}+1\Big).\nonumber
\end{align}
A short calculation shows that $(\tilde{\mathcal{L}}_3+\mathcal{L}_3) = (\tilde{\mathcal{L}}_4+\mathcal{L}_4)$. As they are equivalent we pick $(\tilde{\mathcal{L}}_4+\mathcal{L}_4)$ to participate in our action because there are fewer parity factors. 
For what follows, we define 4 parity factors that will be useful: 
\begin{align}
P_0&\equiv (-1)^{c(b+c)},\\
P_1&\equiv (-1)^{d(e+f+g+a+b+c)+c(e+f+a+b)+b(e+a)+e},\\
P_2&\equiv(-1)^{d(f+g+b+c)+c(f+b)},\\
P_3&\equiv(-1)^{a+f(c+d)+g(d+f)+h(f+g)}\Big((-1)^{a(a+b+c+d)+b(c+d)+cd+f(g+h)+gh}+1\Big).
\end{align}
Reintroducing the scale $\lambda_0$ and introducing two coupling constants, $\alpha_0$ and $\beta_0$, the final form of the super TW gravity action is 
\begin{align}
S_{STW} & = \alpha_0 \int \tensor{\mathcal{K}}{^a_{bcd}} \tensor{\delta}{^c_a} g^{db} P_0 G^{\frac{1}{2}} d \lambda d^{m,n}x \\
& + \beta_0 \int \tensor{\mathcal{K}}{^a_{bcd}} \tensor{\mathcal{K}}{^e_{fgh}} \Big(\tensor{\mathcal{B}}{^{hdgcfb}_{ea}} - \lambda^2_0 g^{hd} g^{gc} g^{fb} g_e g_a P_1 \Big) G^{\frac{1}{2}} d\lambda d^{m,n}x \nonumber \\
& - \beta_0 \lambda^2_0 \int \tensor{\Breve{\mathcal{K}}}{_{bcd}} \tensor{\Breve{\mathcal{K}}}{_{fgh}} g^{hd} g^{gc} g^{fb} P_2 G^{\frac{1}{2}} d\lambda d^{m,n}x \nonumber\\
& - \beta_0 \lambda^2_0 \int \tensor{\mathcal{K}}{^a_{bcd}} \tensor{\Breve{\mathcal{K}}}{_{fgh}} g^{hd} g^{gc} g^{fb} g_a P_3 G^{\frac{1}{2}} d\lambda d^{m,n}x.\nonumber
\end{align}

As stated earlier, setting $\mathcal{P}$ to zero recovers Einstein-Hilbert from the super projective Riemann curvature tensor since
\begin{align}
\tensor{\mathcal{K}}{^a_{bcd}} & = \tensor{R}{^a_{bcd}}, \text{        and} \\
\tensor{\breve{\mathcal{K}}}{_{bcd}} & = - \alpha_a\tensor{R}{^a_{bcd}}.
\end{align}
Define the constant $C$ as, 
\begin{align}
C & = \int_{\lambda_1}^{\lambda_2}\frac{\lambda_0}{\lambda}d\lambda=\lambda_0\text{log}\bigg(\frac{\lambda_2}{\lambda_1}\bigg), 
\end{align}
where $0<\lambda_1<\lambda_2<\infty.$ Recalling the relationship between $G$ and $g$, we have the simplification 
\begin{align}
\int_{\lambda_1}^{\lambda_2} G^{\frac{1}{2}} d \lambda = C  g^{\frac{1}{2}}. \end{align}
Setting $n$ to zero and $m$ to four, and letting $\mathcal{P} \to 0$, the super TW action reduces to 
\begin{align}
{S_{TW}} & = \alpha_0 C \int Rg^{\frac{1}{2}}d^4x + \beta_0 C \int \big( \tensor{R}{^a_{bcd}} \tensor{R}{_a^{bcd}} - 4R_{bd} R^{bd} + R^2 \big) g^{\frac{1}{2}} d^4x. \nonumber
\end{align}
The first term is the Einstein-Hilbert action, while the second term is the Gauss-Bonnet action, which does not contribute to the Einstein field equations in four dimensions \cite{Lanczos:1938sf}.

\section{The Field Equations}

The variation of the canonical measure function is given by 
\begin{align}
\delta g^{\frac{1}{2}}&=\frac{1}{2}g^{-\frac{1}{2}}\delta g=\frac{1}{2}g^{\frac{1}{2}}(-1)^i{^ig^j}\delta{_jg_i}.  \end{align}
The variation of the inverse metric arises from 
\begin{align}
0 & = \delta({^a\delta_c}), \\
\delta g^{ad}&=\delta g^{ab}{_bg_c}g^{cd}=-g^{ab}\delta{_bg_c}g^{cd}=-(-1)^{(b+c)(c+d)}g^{ab}g^{cd}\delta{_bg_c}.
\end{align}
Let $M$ be a Riemannian supermanifold. If $M$ is compact, then \cite{DeWitt:2012mdz}
\begin{align}
\int_M(-1)^i(g^{\frac{1}{2}}X^i)_{,i}d^{m,n}x=\int_M(-1)^ig^{\frac{1}{2}}\tensor{X}{^i_{;i}}d^{m,n}x=0,
\end{align}
implying 
\begin{align}
(\delta\tensor{\Gamma}{^j_{ji}}g^{ki})_{;k}&=(-1)^{k(i+k)}\delta\tensor{\Gamma}{^j_{ji;k}}g^{ki},\\
-g^{\frac{1}{2}}(-1)^j\delta\tensor{\Gamma}{^j_{ji;k}}g^{ki}&=-g^{\frac{1}{2}}(-1)^{j+k(i+k)}(\delta\tensor{\Gamma}{^j_{ji}}g^{ki})_{;k}\nonumber\\
&=-g^{\frac{1}{2}}(-1)^{k}((-1)^j\delta\tensor{\Gamma}{^j_{ji}}g^{ik})_{;k}=-g^{\frac{1}{2}}(-1)^k\tensor{X}{^k_{;k}},\\
X^k&=(-1)^j\delta\tensor{\Gamma}{^j_{ji}}g^{ik},\\
(\delta\tensor{\Gamma}{^j_{ik}}g^{ki})_{;j}&=(-1)^{j(i+k)}\delta\tensor{\Gamma}{^j_{ik;j}}g^{ki},\\
g^{\frac{1}{2}}(-1)^{j(i+j+k)}\delta\tensor{\Gamma}{^j_{ik;j}}g^{ki}&=g^{\frac{1}{2}}(-1)^j(\delta\tensor{\Gamma}{^j_{ik}}g^{ki})_{;j}=g^{\frac{1}{2}}(-1)^j\tensor{Y}{^j_{;j}},\\
Y^j&=\delta\tensor{\Gamma}{^j_{ik}}g^{ki}.
\end{align}

We are now in a position to find field equations of the dynamical fields $\Pi$, $\mathcal{D}$, and $g$ in the action. We express the action in terms of the super projective Cotton-York symbol, reducing the action to a functional that is on $M$, viz, 
\begin{align}
S_{STW} & = \sum_{i=1}^5 S_i,\\
S_1 & = \alpha_0 C \int \tensor{\mathcal{K}}{^a_{bcd}} \tensor{\delta}{^c_a}g^{db}P_0g^{\frac{1}{2}}d^{m,n}x,\\
S_2 & = \beta_0 C \int \tensor{\mathcal{K}}{^a_{bcd}} \tensor{\mathcal{K}}{^e_{fgh}} \tensor{\mathcal{B}}{^{hdgcfb}_{ea}} g^{\frac{1}{2}} d^{m,n}x, \\
S_3 & = - \beta_0 C \lambda^2_0 \int \tensor{\mathcal{K}}{^a_{bcd}}\tensor{\mathcal{K}}{^e_{fgh}}g^{hd}g^{gc}g^{fb}g_eg_aP_1g^{\frac{1}{2}}d^{m,n}x,\\
S_4 & = - \beta_0 C \lambda^2_0 \int \tensor{\breve{\mathcal{K}}}{_{bcd}} \tensor{\breve{\mathcal{K}}}{_{fgh}} g^{hd} g^{gc} g^{fb} P_2 g^{\frac{1}{2}} d^{m,n}x,\\
S_5 & = - \beta_0 C \lambda^2_0 \int \tensor{\mathcal{K}}{^a_{bcd}} \tensor{\breve{\mathcal{K}}}{_{fgh}} g^{hd} g^{gc} g^{fb} g_a P_3 g^{\frac{1}{2}} d^{m,n}x,\\
\tensor{\mathcal{B}}{^{hdgcfb}_{ea}}&=B_1g^{hd}g^{gc}g^{fb}g_{ea}-B_2\tensor{\delta}{^g_e}\tensor{\delta}{^c_a}g^{fb}g^{hd}+B_3\tensor{\delta}{^g_e}\tensor{\delta}{^c_a}g^{hf}
g^{db}. 
\end{align}
Observe that the dependence on $\Pi$ and $\mathcal{D}$ resides in the super projective Cotton-York symbol and the super projective Riemann curvature tensor, while the metric dependence resides only in the superdeterminant of the metric, the inverse metric, and the super Gauss-Bonnet tensor. Before we proceed, we note that $S_{STW}$ can be expressed differently with a particular combination \cite{Brensinger:2020gcv,Brensinger:2017gtb} of the nontrivial coefficients of the super projective Riemann curvature tensor on $VM.$ This combination happens to be a tensor over $M$, known as the projective Cotton-York tensor. The ungraded version is 
\begin{align}
g_a \tensor{K}{^a_{bcd}} + \breve{\mathcal{K}}_{bcd}.   
\end{align}
In every variation below, we hide the constants, $\alpha_0$, $\beta_0$, $\lambda_0$ and $C.$

\subsection{Field Equations for $\Pi $ }
Varying $S_1$ with respect to the super fundamental projective invariant gives
\begin{align}
\delta S_1 = & \int \delta \tensor{\Pi}{^x_{yz}} \bigg(\tensor{\delta}{^a_x} \tensor{\delta}{^y_b} \tensor{\delta}{^z_c}\tensor{{\mathcal{F}_1}}{^c_a^{db}_{,d}}(-1)^{d(a+b+c+d)+c}\\
&-\tensor{\delta}{^a_x}\tensor{\delta}{^y_b}\tensor{\delta}{^z_d}\tensor{{\mathcal{F}_1}}{^c_a^{db}_{,c}}(-1)^{c(a+b)}\nonumber\\
&+\tensor{\delta}{^a_x}\tensor{\delta}{^y_f}\tensor{\delta}{^z_c}\tensor{\Pi}{^f_{bd}}\tensor{{\mathcal{F}_1}}{^c_a^{db}}(-1)^{c(b+c+f)}\nonumber\\
&+\tensor{\Pi}{^a_{fc}}\tensor{\delta}{^f_x}\tensor{\delta}{^y_b}\tensor{\delta}{^z_d}\tensor{{\mathcal{F}_1}}{^c_a^{db}}(-1)^{c(b+c+f)+(x+y+z)(a+c+f)}\nonumber\\
&-\tensor{\delta}{^a_x}\tensor{\delta}{^y_f}\tensor{\delta}{^z_d}\tensor{\Pi}{^f_{bc}}\tensor{{\mathcal{F}_1}}{^c_a^{db}}(-1)^{d(b+c+f)+c}\nonumber\\
&-\tensor{\Pi}{^a_{fd}}\tensor{\delta}{^f_x}\tensor{\delta}{^y_b}\tensor{\delta}{^z_c}\tensor{{\mathcal{F}_1}}{^c_a^{db}}(-1)^{d(b+c+f)+c+(x+y+z)(a+d+f)}\bigg), \nonumber \\ 
\tensor{{\mathcal{F}_1}}{^c_a^{db}} & = (-1)^{bc} \tensor{\delta}{^c_a} g^{db} g^{\frac{1}{2}}, \nonumber 
\end{align}
The variation of $S_2$ with respect to $\Pi$ is
\begin{align}
\delta S_2 & = \int\delta\tensor{\Pi}{^x_{yz}}\bigg(\tensor{\delta}{^a_x}\tensor{\delta}{^y_b}\tensor{\delta}{^z_c}\tensor{{\mathcal{F}_2}}{^{dcb}_{a,d}}(-1)^{d(a+b+c+d)}\\
&-\tensor{\delta}{^a_x}\tensor{\delta}{^y_b}\tensor{\delta}{^z_d}\tensor{{\mathcal{F}_2}}{^{dcb}_{a,c}}(-1)^{c(a+b+c)} + \tensor{\delta}{^a_x}\tensor{\delta}{^y_f}\tensor{\delta}{^z_c}\tensor{\Pi}{^f_{bd}}\tensor{{\mathcal{F}_2}}{^{dcb}_a}(-1)^{c(f+b)}\nonumber\\
    & + \tensor{\delta}{^f_x}\tensor{\delta}{^y_b}\tensor{\delta}{^z_d}\tensor{\Pi}{^a_{fc}}\tensor{{\mathcal{F}_2}}{^{dcb}_a}(-1)^{c(f+b)+(f+b+d)(a+f+c)}\nonumber\\
    & - \tensor{\delta}{^a_x}\tensor{\delta}{^y_f}\tensor{\delta}{^z_d}\tensor{\Pi}{^f_{bc}}\tensor{{\mathcal{F}_2}}{^{dcb}_a}(-1)^{d(b+c+f)}\nonumber\\
    & - \tensor{\delta}{^f_x}\tensor{\delta}{^y_b}\tensor{\delta}{^z_c}\tensor{\Pi}{^a_{fd}}\tensor{{\mathcal{F}_2}}{^{dcb}_a}(-1)^{d(b+c+f)+(a+f+d)(f+b+c)}\bigg) \nonumber \\ 
\tensor{{\mathcal{F}_2}}{^{dcb}_a}&=\tensor{\mathcal{K}}{^e_{fgh}}\bigg(\tensor{\mathcal{B}}{^{hdgcfb}_{ea}}+\tensor{\mathcal{B}}{^{dhcgbf}_{ae}}P_7\bigg)g^{\frac{1}{2}}. \\
P_7&=(-1)^{(a+b+c+d)(e+f+g+h)}.
\end{align}
The variation of $S_3$ with respect to $\Pi$ is 
\begin{align}
\delta S_3&=-\int\delta\tensor{\Pi}{^x_{yz}}\bigg(\tensor{\delta}{^a_x}\tensor{\delta}{^y_b}\tensor{\delta}{^z_c}\tensor{{\mathcal{F}_3}}{^{dcb}_{a,d}}(-1)^{d(a+b+c+d)}\\
&-\tensor{\delta}{^a_x}\tensor{\delta}{^y_b}\tensor{\delta}{^z_d}\tensor{{\mathcal{F}_3}}{^{dcb}_{a,c}}(-1)^{c(a+b+c)}\nonumber\\
&+\tensor{\delta}{^a_x}\tensor{\delta}{^y_f}\tensor{\delta}{^z_c}\tensor{\Pi}{^f_{bd}}\tensor{{\mathcal{F}_3}}{^{dcb}_a}(-1)^{c(f+b)}\nonumber\\
&+\tensor{\delta}{^f_x}\tensor{\delta}{^y_b}\tensor{\delta}{^z_d}\tensor{\Pi}{^a_{fc}}\tensor{{\mathcal{F}_3}}{^{dcb}_a}(-1)^{c(f+b)+(f+b+d)(a+f+c)}\nonumber\\
&-\tensor{\delta}{^a_x}\tensor{\delta}{^y_f}\tensor{\delta}{^z_d}\tensor{\Pi}{^f_{bc}}\tensor{{\mathcal{F}_3}}{^{dcb}_a}(-1)^{d(b+c+f)}\nonumber\\
&-\tensor{\delta}{^f_x}\tensor{\delta}{^y_b}\tensor{\delta}{^z_c}\tensor{\Pi}{^a_{fd}}\tensor{{\mathcal{F}_3}}{^{dcb}_a}(-1)^{d(b+c+f)+(a+f+d)(f+b+c)}\bigg).\nonumber
\end{align}
The variation of $S_4$ with respect to $\Pi$ is 
\begin{align}
\delta S_4 = & - \int\delta\tensor{\Pi}{^x_{yz}}\bigg(\tensor{\delta}{^e_x}\tensor{\delta}{^y_b}\tensor{\delta}{^z_d}\mathcal{D}_{ec}(-1)^{(e+c)(e+b+d)+c(b+e)} \\
& - \tensor{\delta}{^e_x}\tensor{\delta}{^y_b}\tensor{\delta}{^z_c}\mathcal{D}_{ed}(-1)^{d(b+c+e)+(e+d)(e+b+c)}\bigg)\tensor{{\mathcal{F}_4}}{^{dcb}_a}, \nonumber \\ 
\tensor{{\mathcal{F}_4}}{^{dcb}_a} & = \tensor{\mathcal{K}}{_{fgh}} \bigg(g^{hd}g^{gc}g^{fb}P_2+g^{dh}g^{cg}g^{bf}\tilde{P}_2P_8\bigg)g^{\frac{1}{2}},\\
\tilde{P}_2&=(-1)^{h(f+g+b+c)+g(f+b)},\\
P_8&=(-1)^{(b+c+d)(f+g+h)}.
\end{align}
Finally, the variation of $S_5$ with respect to $\Pi$.
\begin{align}
\delta S_5 = & - \int \delta\tensor{\Pi}{^x_{yz}}\bigg(\tensor{\delta}{^a_x}\tensor{\delta}{^y_b}\tensor{\delta}{^z_c}\tensor{{\mathcal{F}_5}}{^{dcb}_{a,d}}(-1)^{d(a+b+c+d)}\\
&-\tensor{\delta}{^a_x}\tensor{\delta}{^y_b}\tensor{\delta}{^z_d}\tensor{{\mathcal{F}_5}}{^{dcb}_{a,c}}(-1)^{c(a+b+c)}\nonumber\\
&+\tensor{\delta}{^a_x}\tensor{\delta}{^y_f}\tensor{\delta}{^z_c}\tensor{\Pi}{^f_{bd}}\tensor{{\mathcal{F}_5}}{^{dcb}_a}(-1)^{c(f+b)}\nonumber\\
&+\tensor{\delta}{^f_x}\tensor{\delta}{^y_b}\tensor{\delta}{^z_d}\tensor{\Pi}{^a_{fc}}\tensor{{\mathcal{F}_5}}{^{dcb}_a}(-1)^{c(f+b)+(f+b+d)(a+f+c)}\nonumber\\
&-\tensor{\delta}{^a_x}\tensor{\delta}{^y_f}\tensor{\delta}{^z_d}\tensor{\Pi}{^f_{bc}}\tensor{{\mathcal{F}_5}}{^{dcb}_a}(-1)^{d(b+c+f)}\nonumber\\
&-\tensor{\delta}{^f_x}\tensor{\delta}{^y_b}\tensor{\delta}{^z_c}\tensor{\Pi}{^a_{fd}}\tensor{{\mathcal{F}_5}}{^{dcb}_a}(-1)^{d(b+c+f)+(a+f+d)(f+b+c)}\nonumber\\
&+\tensor{\delta}{^e_x}\tensor{\delta}{^y_f}\tensor{\delta}{^z_h}\mathcal{D}_{eg}\tensor{{\mathcal{F}_6}}{^{hgf}}(-1)^{(e+g)(e+f+h)+g(f+e)}\nonumber\\
&-\tensor{\delta}{^e_x}\tensor{\delta}{^y_f}\tensor{\delta}{^z_g}\mathcal{D}_{eh}\tensor{{\mathcal{F}_6}}{^{hgf}}(-1)^{h(f+g+e)+(e+h)(e+f+g)}\bigg), \nonumber \\ 
\tensor{{\mathcal{F}_5}}{^{dcb}_a} & = \tensor{\mathcal{K}}{_{fgh}} g^{hd} g^{gc} g^{fb} g_a P_3 g^{\frac{1}{2}},\\
\tensor{{\mathcal{F}_6}}{^{hgf}} & = \tensor{\mathcal{K}}{^a_{bcd}} g^{hd} g^{gc} g^{fb} g_a P_3 P_9 g^{\frac{1}{2}},\\
P_9 & = (-1)^{(a+b+c+d)(f+g+h)}.
\end{align}
Adding up the variations gives the field equations for $\Pi$. The full field equations can be found are in the appendix (see Eq.[\ref{PiEquationofMotion}]).

\subsection{Field Equations for $\mathcal{D}_{bd}$}
Again we start with $S_1.$
\begin{align}
\delta S_1 = & \int \delta \mathcal{D}_{xy} \bigg(\tensor{\delta}{^x_b} \tensor{\delta}{^y_d} \tensor{\delta}{^a_c} (-1)^{bc+(a+c)(b+d)} - \tensor{\delta}{^x_b}\tensor{\delta}{^y_c}\tensor{\delta}{^a_d}(-1)^{d(b+c)+(b+c)(a+d)}\bigg)\tensor{{\mathcal{F}_7}}{^c_a^{db}},\nonumber\\
\tensor{{\mathcal{F}_7}}{^c_a^{db}}&=\tensor{\delta}{^c_a}g^{db}P_0g^{\frac{1}{2}}.
\end{align}
The variation of $S_2$ is 
\begin{align}
\delta S_2 & = \int\delta\mathcal{D}_{xy}\bigg(\tensor{\delta}{^x_b}\tensor{\delta}{^y_d}\tensor{\delta}{^a_c}(-1)^{bc+(a+c)(b+d)} - \tensor{\delta}{^x_b}\tensor{\delta}{^y_c}\tensor{\delta}{^a_d}(-1)^{d(b+c)+(b+c)(a+d)}\bigg)\tensor{{\mathcal{F}_2}}{^{dcb}_a}. \nonumber
\end{align}
The variation of $S_3$ is
\begin{align}
\delta S_3 & = - \int \delta \mathcal{D}_{xy} \bigg( \tensor{\delta}{^x_b} \tensor{\delta}{^y_d} \tensor{\delta}{^a_c} (-1)^{bc+(a+c)(b+d)} - \tensor{\delta}{^x_b}\tensor{\delta}{^y_c}\tensor{\delta}{^a_d}(-1)^{d(b+c)+(b+c)(a+d)}\bigg)\tensor{{\mathcal{F}_3}}{^{dcb}_a}.\nonumber
\end{align}
The variation of $S_4$ is 
\begin{align}
\delta S_4 & = - \int \delta \mathcal{D}_{xy}\bigg(-\tensor{\delta}{^x_b}\tensor{\delta}{^y_d}\tensor{{\mathcal{F}_4}}{^{dcb}_{a,c}}(-1)^{dc+c(a+b+c+d)}\\
&+\tensor{\delta}{^x_b}\tensor{\delta}{^y_c}\tensor{{\mathcal{F}_4}}{^{dcb}_{a,d}}(-1)^{d(a+b+c+d)}\nonumber\\
& + \tensor{\delta}{^x_e}\tensor{\delta}{^y_c}\tensor{\Pi}{^e_{bd}}\tensor{{\mathcal{F}_4}}{^{dcb}_a}(-1)^{c(b+e)} - \tensor{\delta}{^x_e}\tensor{\delta}{^y_d}\tensor{\Pi}{^e_{bc}}\tensor{{\mathcal{F}_4}}{^{dcb}_a}(-1)^{d(b+c+e)}\bigg), \nonumber
\end{align}
and lastly the variation of $S_5$ is 
\begin{align}
\delta S_5 & = -\int\delta\mathcal{D}_{xy}\bigg(\tensor{\delta}{^x_b}\tensor{\delta}{^y_d}\tensor{\delta}{^a_c}\tensor{{\mathcal{F}_5}}{^{dcb}_a}(-1)^{bc+(a+c)(b+d)}\\
&-\tensor{\delta}{^x_b}\tensor{\delta}{^y_c}\tensor{\delta}{^a_d}\tensor{{\mathcal{F}_5}}{^{dcb}_a}(-1)^{d(b+c)+(b+c)(a+d)}\nonumber\\
    & - \tensor{\delta}{^x_f}\tensor{\delta}{^y_h}\tensor{{\mathcal{F}_6}}{^{hgf}_{,g}}(-1)^{hg+g(f+g+h)} + \tensor{\delta}{^x_f}\tensor{\delta}{^y_g}\tensor{{\mathcal{F}_6}}{^{hgf}_{,h}}(-1)^{h(f+g+h)}\nonumber\\
    & + \tensor{\delta}{^x_e}\tensor{\delta}{^y_g}\tensor{\Pi}{^e_{fh}}\tensor{{\mathcal{F}_6}}{^{hgf}}(-1)^{g(f+e)} - \tensor{\delta}{^x_e}\tensor{\delta}{^y_h}\tensor{\Pi}{^e_{fg}}\tensor{{\mathcal{F}_6}}{^{hgf}}(-1)^{h(f+g+e)}\bigg).\nonumber
\end{align}
The full field equations are again in the appendix (see Eq.[\ref{FieldEqDiff}]).

\subsection{The Field Equations for $g_{ab}$}

Let's begin by rearranging and relabeling the action to make the metric dependence explicit, while also keeping in mind the connection and metric are independent. We start by breaking up the action to give 
\begin{align}
S_1 & = \int \sqrt{g}g^{db}\tensor{{\mathcal{F}_8}}{_{bd}},\\
\tensor{{\mathcal{F}_8}}{_{bd}}&=\tensor{\mathcal{K}}{^a_{bcd}}\tensor{\delta}{^c_a}P_0(-1)^{b+d},\\
S_2 & = \int\sqrt{g}\tensor{\mathcal{B}}{^{hdgcfb}_{ea}}\tensor{{\mathcal{F}_9}}{^a_{bcd}^e_{fgh}},\\
\tensor{{\mathcal{F}_9}}{^a_{bcd}^e_{fgh}}&=\tensor{\mathcal{K}}{^a_{bcd}}\tensor{\mathcal{K}}{^e_{fgh}}P_{10},\\
P_{10}&=(-1)^{a+b+c+d+e+f+g+h},\\
S_3 & = \int\sqrt{g} g^{hd} g^{gc} g^{fb} g_e g_a\tensor{{\mathcal{F}_{10}}}{^a_{bcd}^e_{fgh}},\\
\tensor{{\mathcal{F}_{10}}}{^a_{bcd}^e_{fgh}}&=-\tensor{\mathcal{K}}{^a_{bcd}}\tensor{\mathcal{K}}{^e_{fgh}}P_1P_{10},\\
S_4 & = \int\sqrt{g}g^{hd}g^{gc}g^{fb}\tensor{{\mathcal{F}_{11}}}{_{bcdfgh}},\\
\tensor{{\mathcal{F}_{11}}}{_{bcdfgh}}&=-\tensor{\mathcal{K}}{_{bcd}}\tensor{\mathcal{K}}{_{fgh}}P_2(-1)^{b+c+d+f+g+h},\\
S_5 & = \int \sqrt{g} g^{hd}g^{gc}g^{fb}g_a\tensor{{\mathcal{F}_{12}}}{^a_{bcdfgh}},\\
\tensor{{\mathcal{F}_{12}}}{^a_{bcdfgh}}&=-\tensor{\mathcal{K}}{^a_{bcd}}\tensor{\mathcal{K}}{_{fgh}}P_3(-1)^{a+b+c+d+f+g+h}.
\end{align}
We will vary with respect to $g^{ab}$. The family of variations that will be needed are 
\begin{align}
\delta g_c & = \delta g^{ab} \tensor{{V_1}}{_{bac}} + \tensor{{\delta g}}{^{ab}_{,c}} \tensor{{V_2}}{_{ba}} \nonumber \\
\tensor{{V_1}}{_{bac}}&=\frac{1}{2D}g_{ba,c}(-1)^{a+b},\\
\tensor{{V_2}}{_{ba}}&=\frac{1}{2D}g_{ba}(-1)^{c(a+b)+a+b},\\
\tilde{B}_2&=-(-1)^{(f+b+h+d)(g+e+c+a)}B_2,\\
\tilde{B}_3&=(-1)^{(h+f+d+b)(g+e+c+a)}B_3,\\
\delta\tensor{\mathcal{B}}{^{hdgcfb}_{ea}} & = \delta g^{xy}\Bigg[B_1\bigg(\tensor{\delta}{^h_x}\tensor{\delta}{^d_y}g^{gc}g^{fb}g_{ea} + \tensor{\delta}{^g_x} \tensor{\delta}{^c_y}g^{hd}g^{fb}g_{ea}(-1)^{(h+d)(g+c)}\nonumber\\
&+\tensor{\delta}{^f_x}\tensor{\delta}{^b_y}g^{hd}g^{gc}g_{ea}(-1)^{(h+d+g+c)(f+b)}\nonumber\\
&+\tensor{{V_0}}{_{exya}}g^{hd}g^{gc}g^{fb}(-1)^{(h+d+g+c+f+b)(e+a)}\bigg)\nonumber\\
&+\tilde{B}_2\bigg(\tensor{\delta}{^f_x}\tensor{\delta}{^b_y}g^{hd}\tensor{\delta}{^g_e}\tensor{\delta}{^c_a}+\tensor{\delta}{^h_x}\tensor{\delta}{^d_y}g^{fb}\tensor{\delta}{^g_e}\tensor{\delta}{^c_a}(-1)^{(f+b)(h+d)}\bigg)\nonumber\\
&+\tilde{B}_3\bigg(\tensor{\delta}{^h_x}\tensor{\delta}{^f_y}g^{db}\tensor{\delta}{^g_e}\tensor{\delta}{^c_a}+\tensor{\delta}{^d_x}\tensor{\delta}{^b_y}g^{hf}\tensor{\delta}{^g_e}\tensor{\delta}{^c_a}(-1)^{(h+f)(d+b)}\bigg)\Bigg],\nonumber\\
\delta g_a&=\delta g^{xy}\tensor{{V_1}}{_{yxa}}+\tensor{{\delta g}}{^{xy}_{,a}}\tensor{{V_2}}{_{yx}}.
\end{align}
Using the above identities, the variation of $S_1$ with respect to the metric yields 
\begin{align}
\delta S_1 & = \int \delta g^{xy} \bigg[ \bigg(-\frac{1}{2}\sqrt{g}{_yg_x}(-1)^x\bigg)g^{db}\tensor{{\mathcal{F}_8}}{_{bd}}+\tensor{\delta}{^d_x}\tensor{\delta}{^b_y}\sqrt{g}\tensor{{\mathcal{F}_8}}{_{bd}}\bigg].
\end{align}
Similarly, the variation of $S_2$ with respect to the metric is 
\begin{align}
\delta S_2 & = \int \delta g^{xy} \Bigg[ \bigg(-\frac{1}{2}\sqrt{g}{_yg_x}(-1)^x\bigg)\tensor{\mathcal{B}}{^{hdgcfb}_{ea}}\tensor{{\mathcal{F}_9}}{^a_{bcd}^e_{fgh}} \\
    & + \Bigg(B_1 \bigg(\tensor{\delta}{^h_x} \tensor{\delta}{^d_y} g^{gc} g^{fb} g_{ea} + \tensor{\delta}{^g_x} \tensor{\delta}{^c_y} g^{hd} g^{fb} g_{ea} (-1)^{(h+d)(g+c)} \nonumber\\
    & + \tensor{\delta}{^f_x} \tensor{\delta}{^b_y} g^{hd} g^{gc} g_{ea} (-1)^{(h+d+g+c)(f+b)} \nonumber\\
    & + \tensor{{V_0}}{_{exya}} g^{hd} g^{gc} g^{fb} (-1)^{(h+d+g+c+f+b)(e+a)} \bigg) \nonumber\\
    & + \tilde{B}_2 \bigg(\tensor{\delta}{^f_x} \tensor{\delta}{^b_y} g^{hd} \tensor{\delta}{^g_e} \tensor{\delta}{^c_a} + \tensor{\delta}{^h_x} \tensor{\delta}{^d_y} g^{fb} \tensor{\delta}{^g_e} \tensor{\delta}{^c_a} (-1)^{(f+b)(h+d)} \bigg) \nonumber\\
    & + \tilde{B}_3 \bigg( \tensor{\delta}{^h_x} \tensor{\delta}{^f_y} g^{db} \tensor{\delta}{^g_e} \tensor{\delta}{^c_a} + \tensor{\delta}{^d_x} \tensor{\delta}{^b_y} g^{hf} \tensor{\delta}{^g_e} \tensor{\delta}{^c_a} (-1)^{(h+f)(d+b)} \bigg)\Bigg) \sqrt{g} \tensor{{\mathcal{F}_9}}{^a_{bcd}^e_{fgh}} \Bigg]. \nonumber
\end{align}
First, we define $\mathcal{F}_{13}$ and $\mathcal{F}_{14}$ to clean up the variation below.
\begin{align}
\tensor{{\mathcal{F}_{13}}}{^{ae}}&=\sqrt{g}g^{hd}g^{gc}g^{fb}\tensor{{\mathcal{F}_{10}}}{^a_{bcd}^e_{fgh}}(-1)^{(e+a)(h+d+g+c+f+b)},\\
\tensor{{\mathcal{F}_{14}}}{_{bcdfgh}}&=g_eg_a\tensor{{\mathcal{F}_{10}}}{^a_{bcd}^e_{fgh}}.
\end{align}
The variation of $S_3$ with respect to the metric.
\begin{align}
\delta S_3 & = \int \delta g^{xy} \Bigg[\tensor{{V_1}}{_{yxe}} g_a \tensor{{\mathcal{F}_{13}}}{^{ae}} - \Big(\tensor{{V_2}}{_{yx}} g_a \tensor{{\mathcal{F}_{13}}}{^{ae}}\Big)_{,e} \\
& + \tensor{{V_1}}{_{yxa}} g_e \tensor{{\mathcal{F}_{13}}}{^{ae}}(-1)^{ae} - \Big(\tensor{{V_2}}{_{yx}} g_e \tensor{{\mathcal{F}_{13}}}{^{ae}} (-1)^{ae} \Big)_{,a} \nonumber\\
& + \Bigg( \bigg( - \frac{1}{2} \sqrt{g}{_yg_x} (-1)^x \bigg) g^{hd} g^{gc} g^{fb} \nonumber \\
    & + \tensor{\delta}{^h_x} \tensor{\delta}{^d_y} \sqrt{g}g^{gc}g^{fb} + \tensor{\delta}{^g_x} \tensor{\delta}{^c_y} \sqrt{g} g^{hd} g^{fb} (-1)^{(h+d)(g+c)} \nonumber\\
    & + \tensor{\delta}{^f_x} \tensor{\delta}{^b_y} \sqrt{g} g^{hd} g^{gc} (-1)^{(f+b)(h+d+g+c)} \Bigg) \tensor{{\mathcal{F}_{14}}}{_{bcdfgh}} \Bigg]. \nonumber 
\end{align}
The variation of $S_4$ with respect to the metric.
\begin{align}
\delta S_4 & = \int \delta g^{xy}\Bigg[\bigg(-\frac{1}{2}\sqrt{g}{_yg_x}(-1)^x\bigg)g^{hd}g^{gc}g^{fb}\\
    & + \tensor{\delta}{^h_x}\tensor{\delta}{^d_y}\sqrt{g}g^{gc}g^{fb} + \tensor{\delta}{^g_x}\tensor{\delta}{^c_y}\sqrt{g}g^{hd}g^{fb}(-1)^{(h+d)(g+c)} \nonumber\\
    & + \tensor{\delta}{^f_x}\tensor{\delta}{^b_y}\sqrt{g}g^{hd}g^{gc}(-1)^{(f+b)(h+d+g+c)}\Bigg]\tensor{{\mathcal{F}_{11}}}{_{bcdfgh}}. \nonumber
\end{align}
We define $\mathcal{F}_{15}$ and $\mathcal{F}_{16}$ in order to clean up the variation below.
\begin{align}
\tensor{{\mathcal{F}_{15}}}{_{bcdfgh}}&=g_a\tensor{{\mathcal{F}_{12}}}{^a_{bcdfgh}}\\
\tensor{{\mathcal{F}_{16}}}{^a}&=\sqrt{g}g^{hd}g^{gc}g^{fb}\tensor{{\mathcal{F}_{12}}}{^a_{bcdfgh}}(-1)^{a(h+d+g+c+f+c)}.
\end{align}
The variation of $S_5$ with respect to the metric.
\begin{align}
\delta S_5 & =\int \delta g^{xy} \Bigg[ \Bigg( \bigg( - \frac{1}{2} \sqrt{g}{_yg_x} (-1)^x\bigg) g^{hd} g^{gc} g^{fb} \\
& + \tensor{\delta}{^h_x} \tensor{\delta}{^d_y} \sqrt{g} g^{gc} g^{fb} +\tensor{\delta}{^g_x}\tensor{\delta}{^c_y}\sqrt{g}g^{hd}g^{fb}(-1)^{(h+d)(g+c)}\nonumber\\
    & + \tensor{\delta}{^f_x}\tensor{\delta}{^b_y}\sqrt{g}g^{hd}g^{gc}(-1)^{(f+b)(h+d+g+c)}\Bigg)\tensor{{\mathcal{F}_{15}}}{_{bcdfgh}}\nonumber\\
    & + \tensor{{V_1}}{_{yxa}}\tensor{{\mathcal{F}_{16}}}{^a}-\bigg(\tensor{{V_2}}{_{yx}}\tensor{{\mathcal{F}_{16}}}{^a}\bigg)_{,a}(-1)^{a(a+x+y)}\Bigg].\nonumber
\end{align}
As before, the equations of motion for $g_{ab}$ are found in the appendix (see Eq.[\ref{FieldEqMetric}]).  From here one may define an energy-momentum tensor.  In order to write the usual Einstein equations, one would be obliged to decompose the fundamental projective invariant into an affine connection and a traceless Palatini tensor. Then one could have the usual Riemannian geometric objects on the left-hand side of these field equations, while the Palatini field equations and the contributions from the diff field would move to the right-hand side forming projective geometric sources. 

\section{Conclusion}

In this note, we have generalized TW gravity to a graded setting in the framework of a DeWitt supermanifold. The super TW gravity action is invariant under super projective transformations, yields second-order partial differential equations for the metric, the super fundamental projective invariant and the super Diffeomorphism field. Our construction generated an infinitesimal coordinate transformation law for the super Diffeomorphism field, which recovered the coadjoint action on a coadjoint Virasoro element in a particular limit. Additionally, setting the number of fermionic coordinates to zero, the number of bosonic coordinates to four, and the tensorial relative to the super Diffeomorphism field to zero, the super TW action simplified to the Einstein-Hilbert action. The super TW action is the natural precursor to understanding a theory of projective super gravity with dynamical projective connections intimately connected to the super Virasoro algebra.  For instance, setting one of indices of the super Diffeomorphism field to a bosonic index and the other to a fermionic index, one would expect the appearance of a spin-3/2 Rarita-Schwinger field. Also, we expect a  supersymmetric version of the super TW action to make contact with the supersymmetric extension of the 2D Polyakov Action \cite{Polyakov:1987zb,Delius:1990pt,Aoyama:1989pw,Bershadsky:1989tc}. The super TW action described in this note is not restricted to supersymmetric coordinates and can be used to investigate other super space phenomenon. Also, our analysis focused completely on tensors and does not address the study of spinors in superspaces and their coupling to the TW connection. Details on the investigation of fermions in TW gravity in the ungraded setting were discussed in detail in \cite{Brensinger:2020gcv}.

\section{Acknowledgement}
The authors thank the members of the Diffeomorphism and Geometry group and the High Energy Phyisics Group at the University of Iowa for constructive dialog throughout this work. CM and PV would like to thank Charles Frohman for useful conversations.

\appendix

\section{Summary of Conventions}

Parity Factors
\begin{align}
P_0&\equiv (-1)^{c(b+c)},\\
P_1&\equiv (-1)^{d(e+f+g+a+b+c)+c(e+f+a+b)+b(e+a)+e},\\
\tilde{P}_1&\equiv(-1)^{h(e+f+g+a+b+c)+g(e+f+a+b)+f(e+a)+a},\\
P_2&\equiv(-1)^{d(f+g+b+c)+c(f+b)},\\
\tilde{P}_2&\equiv(-1)^{h(f+g+b+c)+g(f+b)},\\
P_3&\equiv(-1)^{a+f(c+d)+g(d+f)+h(f+g)}\Big((-1)^{a(a+b+c+d)+b(c+d)+cd+f(g+h)+gh}+1\Big),\\
P_4&\equiv(-1)^{a+e(b+c+d+e)+f(c+d)+g(d+f)+h(f+g)}\big(1-(-1)^{b(c+d)+cd+f(g+h)+gh}\big),\\
P_5&\equiv(-1)^{f(c+d)+g(d+f)+h(f+g)},\\
P_6&\equiv(-1)^{b(c+d)+cd+f(g+h)+gh},\\
P_7&\equiv(-1)^{(a+b+c+d)(e+f+g+h)},\\
P_8&\equiv(-1)^{(b+c+d)(f+g+h)},\\
P_9&\equiv(-1)^{(a+b+c+d)(f+g+h)},\\
P_{10}&\equiv(-1)^{a+b+c+d+e+f+g+h}, \\
B_1&\equiv(-1)^{d(e+f+g+a+b+c)+c(e+f+a+b)+b(e+a)+e},\\
B_2&\equiv4(-1)^{c(f+h+b+c)+a(f+h)+g(f+g)+h(f+b)},\\
\tilde{B}_2&\equiv-(-1)^{(f+b+h+d)(g+e+c+a)}B_2,\\
B_3&\equiv(-1)^{c(b+c)+g(f+g)+(f+h)(a+c)},\\
\tilde{B}_3&\equiv(-1)^{(h+f+d+b)(g+e+c+a)}B_3.
\end{align}

Tensors and Symbols 
\begin{align}
\tensor{{\mathcal{F}_1}}{^c_a^{db}} & = (-1)^{bc} \tensor{\delta}{^c_a} g^{db}g^{\frac{1}{2}},\\
\tensor{{\mathcal{F}_2}}{^{dcb}_a} & = \tensor{\mathcal{K}}{^e_{fgh}}\bigg(\tensor{\mathcal{B}}{^{hdgcfb}_{ea}}+\tensor{\mathcal{B}}{^{dhcgbf}_{ae}}P_7\bigg)g^{\frac{1}{2}},\\
\tensor{{\mathcal{F}_3}}{^{dcb}_a}&=\tensor{\mathcal{K}}{^e_{fgh}}\bigg(g^{hd}g^{gc}g^{fb}g_eg_aP_1+g^{dh}g^{cg}g^{bf}g_ag_e\tilde{P}_1P_7\bigg)g^{\frac{1}{2}},\\
\tensor{{\mathcal{F}_4}}{^{dcb}_a}&=\tensor{\mathcal{K}}{_{fgh}} \bigg(g^{hd} g^{gc} g^{fb}P_2 + g^{dh}g^{cg}g^{bf} \tilde{P}_2P_8\bigg) g^{\frac{1}{2}},\\
\tensor{{\mathcal{F}_5}}{^{dcb}_a}&=\tensor{\mathcal{K}}{_{fgh}}g^{hd}g^{gc}g^{fb}g_aP_3g^{\frac{1}{2}},\\
\tensor{{\mathcal{F}_6}}{^{hgf}}&=\tensor{\mathcal{K}}{^a_{bcd}}g^{hd}g^{gc}g^{fb}g_aP_3P_9g^{\frac{1}{2}},\\
\tensor{{\mathcal{F}_7}}{^c_a^{db}}&=\tensor{\delta}{^c_a}g^{db}P_0g^{\frac{1}{2}},\\
\tensor{{\mathcal{F}_8}}{_{bd}}&=\tensor{\mathcal{K}}{^a_{bcd}}\tensor{\delta}{^c_a}P_0(-1)^{b+d},\\
\tensor{{\mathcal{F}_9}}{^a_{bcd}^e_{fgh}}&=\tensor{\mathcal{K}}{^a_{bcd}}\tensor{\mathcal{K}}{^e_{fgh}}P_{10},\\
\tensor{{\mathcal{F}_{10}}}{^a_{bcd}^e_{fgh}}&=-\tensor{\mathcal{K}}{^a_{bcd}}\tensor{\mathcal{K}}{^e_{fgh}}P_1P_{10},\\
\tensor{{\mathcal{F}_{11}}}{_{bcdfgh}}&=-\tensor{\mathcal{K}}{_{bcd}}\tensor{\mathcal{K}}{_{fgh}}P_2(-1)^{b+c+d+f+g+h},\\
\tensor{{\mathcal{F}_{12}}}{^a_{bcdfgh}}&=-\tensor{\mathcal{K}}{^a_{bcd}}\tensor{\mathcal{K}}{_{fgh}}P_3(-1)^{a+b+c+d+f+g+h},\\
\tensor{{\mathcal{F}_{13}}}{^{ae}}&=\sqrt{g}g^{hd}g^{gc}g^{fb}\tensor{{\mathcal{F}_{10}}}{^a_{bcd}^e_{fgh}}(-1)^{(e+a)(h+d+g+c+f+b)},\\
\tensor{{\mathcal{F}_{14}}}{_{bcdfgh}}&=g_eg_a\tensor{{\mathcal{F}_{10}}}{^a_{bcd}^e_{fgh}},\\
\tensor{{\mathcal{F}_{15}}}{_{bcdfgh}}&=g_a\tensor{{\mathcal{F}_{12}}}{^a_{bcdfgh}},\\
\tensor{{\mathcal{F}_{16}}}{^a}&=\sqrt{g}g^{hd}g^{gc}g^{fb}\tensor{{\mathcal{F}_{12}}}{^a_{bcdfgh}}(-1)^{a(h+d+g+c+f+c)}, \\
\tensor{{V_0}}{_{dabc}}&=-g_{da}g_{bc}(-1)^{(a+b)(a+d)+b},\\
\tensor{{V_1}}{_{bac}}&=\frac{1}{2D}g_{ba,c}(-1)^{a+b},\\
\tensor{{V_2}}{_{ba}}&=\frac{1}{2D}g_{ba}(-1)^{c(a+b)+a+b}.
\end{align}

\section{Field Equations}

\subsection{Field Equations for $\Pi$}

\begin{align}
& \alpha_0 C \bigg( \tensor{\delta}{^a_x} \tensor{\delta}{^y_b} \tensor{\delta}{^z_c} \tensor{{\mathcal{F}_1}}{^c_a^{db}_{,d}} (-1)^{d(a+b+c+d)+c} \label{PiEquationofMotion} \\
    & \qquad - \tensor{\delta}{^a_x} \tensor{\delta}{^y_b} \tensor{\delta}{^z_d} \tensor{{\mathcal{F}_1}}{^c_a^{db}_{,c}} (-1)^{c(a+b)} \nonumber\\
    & \qquad + \tensor{\delta}{^a_x} \tensor{\delta}{^y_f} \tensor{\delta}{^z_c} \tensor{\Pi}{^f_{bd}} \tensor{{\mathcal{F}_1}}{^c_a^{db}} (-1)^{c(b+c+f)}\nonumber\\
    & \qquad + \tensor{\Pi}{^a_{fc}} \tensor{\delta}{^f_x} \tensor{\delta}{^y_b} \tensor{\delta}{^z_d} \tensor{{\mathcal{F}_1}}{^c_a^{db}} (-1)^{c(b+c+f)+(x+y+z)(a+c+f)} \nonumber \\
    & \qquad - \tensor{\delta}{^a_x} \tensor{\delta}{^y_f} \tensor{\delta}{^z_d} \tensor{\Pi}{^f_{bc}} \tensor{{\mathcal{F}_1}}{^c_a^{db}} (-1)^{d(b+c+f)+c} \nonumber \\
    & \qquad - \tensor{\Pi}{^a_{fd}} \tensor{\delta}{^f_x} \tensor{\delta}{^y_b} \tensor{\delta}{^z_c} \tensor{{\mathcal{F}_1}}{^c_a^{db}} (-1)^{d(b+c+f)+c+(x+y+z)(a+d+f)} \bigg) \nonumber \\
& + \beta_0 C \bigg(\tensor{\delta}{^a_x} \tensor{\delta}{^y_b} \tensor{\delta}{^z_c} \tensor{{\mathcal{F}_2}}{^{dcb}_{a,d}} (-1)^{d(a+b+c+d)} \nonumber \\
    & \qquad - \tensor{\delta}{^a_x} \tensor{\delta}{^y_b} \tensor{\delta}{^z_d} \tensor{{\mathcal{F}_2}}{^{dcb}_{a,c}} (-1)^{c(a+b+c)} + \tensor{\delta}{^a_x} \tensor{\delta}{^y_f} \tensor{\delta}{^z_c} \tensor{\Pi}{^f_{bd}} \tensor{{\mathcal{F}_2}}{^{dcb}_a} (-1)^{c(f+b)} \nonumber\\
    & \qquad + \tensor{\delta}{^f_x}\tensor{\delta}{^y_b}\tensor{\delta}{^z_d}\tensor{\Pi}{^a_{fc}}\tensor{{\mathcal{F}_2}}{^{dcb}_a}(-1)^{c(f+b)+(f+b+d)(a+f+c)} \nonumber \\
    & \qquad - \tensor{\delta}{^a_x} \tensor{\delta}{^y_f} \tensor{\delta}{^z_d} \tensor{\Pi}{^f_{bc}} \tensor{{\mathcal{F}_2}}{^{dcb}_a} (-1)^{d(b+c+f)} \nonumber\\
    & \qquad - \tensor{\delta}{^f_x}\tensor{\delta}{^y_b}\tensor{\delta}{^z_c}\tensor{\Pi}{^a_{fd}}\tensor{{\mathcal{F}_2}}{^{dcb}_a}(-1)^{d(b+c+f)+(a+f+d)(f+b+c)}\bigg) \nonumber \\ 
& + \beta_0 C \lambda_0^2 \bigg(\tensor{\delta}{^a_x}\tensor{\delta}{^y_b}\tensor{\delta}{^z_c}\tensor{{\mathcal{F}_3}}{^{dcb}_{a,d}}(-1)^{d(a+b+c+d)} \nonumber \\
    & \qquad - \tensor{\delta}{^a_x} \tensor{\delta}{^y_b} \tensor{\delta}{^z_d} \tensor{{\mathcal{F}_3}}{^{dcb}_{a,c}} (-1)^{c(a+b+c)} \nonumber \\
    & \qquad + \tensor{\delta}{^a_x} \tensor{\delta}{^y_f} \tensor{\delta}{^z_c} \tensor{\Pi}{^f_{bd}} \tensor{{\mathcal{F}_3}}{^{dcb}_a}(-1)^{c(f+b)} \nonumber \\
    & \qquad + \tensor{\delta}{^f_x} \tensor{\delta}{^y_b} \tensor{\delta}{^z_d} \tensor{\Pi}{^a_{fc}} \tensor{{\mathcal{F}_3}}{^{dcb}_a} (-1)^{c(f+b)+(f+b+d)(a+f+c)} \nonumber \\
    & \qquad -\tensor{\delta}{^a_x}\tensor{\delta}{^y_f}\tensor{\delta}{^z_d}\tensor{\Pi}{^f_{bc}}\tensor{{\mathcal{F}_3}}{^{dcb}_a}(-1)^{d(b+c+f)} \nonumber \\
    & \qquad - \tensor{\delta}{^f_x} \tensor{\delta}{^y_b} \tensor{\delta}{^z_c} \tensor{\Pi}{^a_{fd}} \tensor{{\mathcal{F}_3}}{^{dcb}_a} (-1)^{d(b+c+f)+(a+f+d)(f+b+c)} \nonumber \\
& + \tensor{\delta}{^e_x}\tensor{\delta}{^y_b}\tensor{\delta}{^z_d} \mathcal{D}_{ec} \tensor{{\mathcal{F}_4}}{^{dcb}_a} (-1)^{(e+c)(e+b+d)+c(b+e)} \nonumber \\
    & \qquad - \tensor{\delta}{^e_x} \tensor{\delta}{^y_b} \tensor{\delta}{^z_c} \mathcal{D}_{ed} \tensor{{\mathcal{F}_4}}{^{dcb}_a} (-1)^{d(b+c+e)+(e+d)(e+b+c)} \nonumber \\ 
& + \tensor{\delta}{^a_x} \tensor{\delta}{^y_b} \tensor{\delta}{^z_c} \tensor{{\mathcal{F}_5}}{^{dcb}_{a,d}} (-1)^{d(a+b+c+d)} - \tensor{\delta}{^a_x} \tensor{\delta}{^y_b} \tensor{\delta}{^z_d} \tensor{{\mathcal{F}_5}}{^{dcb}_{a,c}} (-1)^{c(a+b+c)} \nonumber \\
    & \qquad + \tensor{\delta}{^a_x} \tensor{\delta}{^y_f} \tensor{\delta}{^z_c} \tensor{\Pi}{^f_{bd}} \tensor{{\mathcal{F}_5}}{^{dcb}_a} (-1)^{c(f+b)} \nonumber - \tensor{\delta}{^a_x} \tensor{\delta}{^y_f} \tensor{\delta}{^z_d} \tensor{\Pi}{^f_{bc}} \tensor{{\mathcal{F}_5}}{^{dcb}_a} (-1)^{d(b+c+f)} \\
    & \qquad + \tensor{\delta}{^f_x} \tensor{\delta}{^y_b} \tensor{\delta}{^z_d} \tensor{\Pi}{^a_{fc}} \tensor{{\mathcal{F}_5}}{^{dcb}_a}(-1)^{c(f+b)+(f+b+d)(a+f+c)} \nonumber \\
    & \qquad -\tensor{\delta}{^f_x}\tensor{\delta}{^y_b}\tensor{\delta}{^z_c}\tensor{\Pi}{^a_{fd}}\tensor{{\mathcal{F}_5}}{^{dcb}_a}(-1)^{d(b+c+f)+(a+f+d)(f+b+c)}\nonumber\\
    & \qquad +\tensor{\delta}{^e_x}\tensor{\delta}{^y_f}\tensor{\delta}{^z_h}\mathcal{D}_{eg}\tensor{{\mathcal{F}_6}}{^{hgf}}(-1)^{(e+g)(e+f+h)+g(f+e)}\nonumber\\
    & \qquad -\tensor{\delta}{^e_x}\tensor{\delta}{^y_f}\tensor{\delta}{^z_g}\mathcal{D}_{eh}\tensor{{\mathcal{F}_6}}{^{hgf}}(-1)^{h(f+g+e)+(e+h)(e+f+g)}\bigg) = 0. \nonumber
\end{align}

\subsection{Field Equations for $\mathcal{D}$}

\begin{align}
& \alpha_0 C \bigg(\tensor{\delta}{^x_b} \tensor{\delta}{^y_d} \tensor{\delta}{^a_c} (-1)^{bc+(a+c)(b+d)} - \tensor{\delta}{^x_b} \tensor{\delta}{^y_c} \tensor{\delta}{^a_d} (-1)^{d(b+c)+(b+c)(a+d)} \bigg) \tensor{{\mathcal{F}_7}}{^c_a^{db}} \label{FieldEqDiff} \\
& + \beta_0 C \bigg( \tensor{\delta}{^x_b} \tensor{\delta}{^y_d} \tensor{\delta}{^a_c} (-1)^{bc+(a+c)(b+d)} - \tensor{\delta}{^x_b} \tensor{\delta}{^y_c} \tensor{\delta}{^a_d} (-1)^{d(b+c)+(b+c)(a+d)} \bigg) \tensor{{\mathcal{F}_2}}{^{dcb}_a} \nonumber \\ 
& + \beta_0 C \lambda_0^2 \bigg( \tensor{\delta}{^x_b} \tensor{\delta}{^y_d} \tensor{\delta}{^a_c} \tensor{{\mathcal{F}_3}}{^{dcb}_a} (-1)^{bc+(a+c)(b+d)} - \tensor{\delta}{^x_b} \tensor{\delta}{^y_c} \tensor{\delta}{^a_d} \tensor{{\mathcal{F}_3}}{^{dcb}_a} (-1)^{d(b+c)+(b+c)(a+d)} \nonumber \\ 
    & \qquad - \tensor{\delta}{^x_b} \tensor{\delta}{^y_d} \tensor{{\mathcal{F}_4}}{^{dcb}_{a,c}} (-1)^{dc+c(a+b+c+d)} + \tensor{\delta}{^x_b} \tensor{\delta}{^y_c} \tensor{{\mathcal{F}_4}}{^{dcb}_{a,d}} (-1)^{d(a+b+c+d)} \nonumber \\
    & \qquad + \tensor{\delta}{^x_e} \tensor{\delta}{^y_c} \tensor{\Pi}{^e_{bd}} \tensor{{\mathcal{F}_4}}{^{dcb}_a}(-1)^{c(b+e)} - \tensor{\delta}{^x_e} \tensor{\delta}{^y_d} \tensor{\Pi}{^e_{bc}} \tensor{{\mathcal{F}_4}}{^{dcb}_a} (-1)^{d(b+c+e)} \nonumber \\ 
    & \qquad + \tensor{\delta}{^x_b}\tensor{\delta}{^y_d}\tensor{\delta}{^a_c}\tensor{{\mathcal{F}_5}}{^{dcb}_a}(-1)^{bc+(a+c)(b+d)} - \tensor{\delta}{^x_b} \tensor{\delta}{^y_c} \tensor{\delta}{^a_d} \tensor{{\mathcal{F}_5}}{^{dcb}_a} (-1)^{d(b+c)+(b+c)(a+d)}\nonumber\\
    & \qquad - \tensor{\delta}{^x_f} \tensor{\delta}{^y_h} \tensor{{\mathcal{F}_6}}{^{hgf}_{,g}}(-1)^{hg+g(f+g+h)} + \tensor{\delta}{^x_f} \tensor{\delta}{^y_g} \tensor{{\mathcal{F}_6}}{^{hgf}_{,h}} (-1)^{h(f+g+h)}\nonumber\\
    & \qquad + \tensor{\delta}{^x_e} \tensor{\delta}{^y_g} \tensor{\Pi}{^e_{fh}} \tensor{{\mathcal{F}_6}}{^{hgf}}(-1)^{g(f+e)} - \tensor{\delta}{^x_e} \tensor{\delta}{^y_h} \tensor{\Pi}{^e_{fg}} \tensor{{\mathcal{F}_6}}{^{hgf}} (-1)^{h(f+g+e)} \bigg) = 0. \nonumber
\end{align}

\subsection{Field Equations for g}

\begin{align}
& \alpha_0 C \bigg( - \frac{1}{2}\sqrt{g} \ {_yg_x} g^{db} \tensor{{\mathcal{F}_8}}{_{bd}} (-1)^x +\tensor{\delta}{^d_x}\tensor{\delta}{^b_y} \sqrt{g} \tensor{{\mathcal{F}_8}}{_{bd}} \bigg) \label{FieldEqMetric} \\ 
& + \beta_0 C \Bigg[ - \frac{1}{2} \sqrt{g} \ {_yg_x} \tensor{\mathcal{B}}{^{hdgcfb}_{ea}} \tensor{{\mathcal{F}_9}}{^a_{bcd}^e_{fgh}} (-1)^x \nonumber \\
    & \qquad + \Bigg(B_1 \bigg(\tensor{\delta}{^h_x} \tensor{\delta}{^d_y} g^{gc} g^{fb} g_{ea} + \tensor{\delta}{^g_x} \tensor{\delta}{^c_y} g^{hd} g^{fb} g_{ea} (-1)^{(h+d)(g+c)} \nonumber\\
    & \qquad + \tensor{\delta}{^f_x} \tensor{\delta}{^b_y} g^{hd} g^{gc} g_{ea} (-1)^{(h+d+g+c)(f+b)} \nonumber\\
    & \qquad + \tensor{{V_0}}{_{exya}} g^{hd} g^{gc} g^{fb} (-1)^{(h+d+g+c+f+b)(e+a)} \bigg) \nonumber\\
    & + \tilde{B}_2 \tensor{\delta}{^f_x} \tensor{\delta}{^b_y} g^{hd} \tensor{\delta}{^g_e} \tensor{\delta}{^c_a} + \tilde{B}_2 \tensor{\delta}{^h_x} \tensor{\delta}{^d_y} g^{fb} \tensor{\delta}{^g_e} \tensor{\delta}{^c_a} (-1)^{(f+b)(h+d)} \nonumber\\
    & \qquad + \tilde{B}_3 \tensor{\delta}{^h_x} \tensor{\delta}{^f_y} g^{db} \tensor{\delta}{^g_e} \tensor{\delta}{^c_a} + \tilde{B}_3 \tensor{\delta}{^d_x} \tensor{\delta}{^b_y} g^{hf} \tensor{\delta}{^g_e} \tensor{\delta}{^c_a} (-1)^{(h+f)(d+b)} \Bigg) \sqrt{g} \tensor{{\mathcal{F}_9}}{^a_{bcd}^e_{fgh}} \Bigg] \nonumber \\ 
& - \beta_0 C \lambda_0^2 \Bigg[\tensor{{V_1}}{_{yxe}} g_a \tensor{{\mathcal{F}_{13}}}{^{ae}} - \Big(\tensor{{V_2}}{_{yx}} g_a \tensor{{\mathcal{F}_{13}}}{^{ae}}\Big)_{,e} \nonumber \\
    & \qquad + \tensor{{V_1}}{_{yxa}} g_e \tensor{{\mathcal{F}_{13}}}{^{ae}}(-1)^{ae} - \Big(\tensor{{V_2}}{_{yx}} g_e \tensor{{\mathcal{F}_{13}}}{^{ae}} (-1)^{ae} \Big)_{,a} \nonumber \\
    & \qquad + \Bigg( - \frac{1}{2} \sqrt{g} \ {_yg_x} g^{hd} g^{gc} g^{fb} (-1)^x \nonumber \\
    & \qquad + \tensor{\delta}{^h_x} \tensor{\delta}{^d_y} \sqrt{g}g^{gc}g^{fb} + \tensor{\delta}{^g_x} \tensor{\delta}{^c_y} \sqrt{g} g^{hd} g^{fb} (-1)^{(h+d)(g+c)} \nonumber\\
    & \qquad + \tensor{\delta}{^f_x} \tensor{\delta}{^b_y} \sqrt{g} g^{hd} g^{gc} (-1)^{(f+b)(h+d+g+c)} \Bigg) \tensor{{\mathcal{F}_{14}}}{_{bcdfgh}} \nonumber \\ 
& + \bigg( - \frac{1}{2} \sqrt{g} \ {_yg_x} g^{hd} g^{gc} g^{fb} (-1)^x \nonumber \\
    & \qquad + \tensor{\delta}{^h_x}\tensor{\delta}{^d_y}\sqrt{g}g^{gc}g^{fb} + \tensor{\delta}{^g_x}\tensor{\delta}{^c_y}\sqrt{g}g^{hd}g^{fb}(-1)^{(h+d)(g+c)} \nonumber\\
    & \qquad + \tensor{\delta}{^f_x}\tensor{\delta}{^b_y}\sqrt{g}g^{hd}g^{gc}(-1)^{(f+b)(h+d+g+c)} \bigg) \tensor{{\mathcal{F}_{11}}}{_{bcdfgh}} \nonumber \\ 
& + \Bigg( - \frac{1}{2} \sqrt{g} \ {_yg_x} g^{hd} g^{gc} g^{fb} (-1)^x \nonumber \\
    & \qquad + \tensor{\delta}{^h_x} \tensor{\delta}{^d_y} \sqrt{g} g^{gc} g^{fb} +\tensor{\delta}{^g_x}\tensor{\delta}{^c_y}\sqrt{g}g^{hd}g^{fb}(-1)^{(h+d)(g+c)}\nonumber\\
    & \qquad + \tensor{\delta}{^f_x}\tensor{\delta}{^b_y}\sqrt{g}g^{hd}g^{gc}(-1)^{(f+b)(h+d+g+c)}\Bigg)\tensor{{\mathcal{F}_{15}}}{_{bcdfgh}}\nonumber\\
    & \qquad + \tensor{{V_1}}{_{yxa}}\tensor{{\mathcal{F}_{16}}}{^a}-\bigg(\tensor{{V_2}}{_{yx}}\tensor{{\mathcal{F}_{16}}}{^a}\bigg)_{,a}(-1)^{a(a+x+y)}\Bigg] = 0. \nonumber
\end{align}

\bibliographystyle{apsrev4-1}
\bibliography{SuperBibliography}
\end{document}